\documentclass[paper]{geophysics}

\usepackage{amssymb}
\usepackage{amsmath}
\usepackage{amscd}
\usepackage{dsfont}
\usepackage{graphicx}
\usepackage{graphics}
\usepackage[noadjust]{cite}
\usepackage{amsthm}
\usepackage{caption}
\usepackage{multirow}
\usepackage{array}
\usepackage{makecell}
\usepackage[norelsize,boxruled]{algorithm2e}
\usepackage{xcolor}
\usepackage{color}
\usepackage{indentfirst}

\newlength{\Oldarrayrulewidth}
\newcommand{\Cline}[2]{%
  \noalign{\global\setlength{\Oldarrayrulewidth}{\arrayrulewidth}}%
  \noalign{\global\setlength{\arrayrulewidth}{#1}}\cline{#2}%
  \noalign{\global\setlength{\arrayrulewidth}{\Oldarrayrulewidth}}}

\renewcommand{\thealgocf}{}

\DeclareMathSizes{10.95}{10}{7}{5}

\newcommand\norm[1]{\left\lVert#1\right\rVert}
\makeatletter
\renewcommand*\env@matrix[1][\arraystretch]{%
  \edef\arraystretch{#1}%
  \hskip -\arraycolsep
  \let\@ifnextchar\new@ifnextchar
  \array{*\c@MaxMatrixCols c}}
\makeatletter

\begin{document}
\title{From constant to variable density inverse extended Born modelling}


\author{Milad Farshad $^{1}$,  Herv\'e Chauris $^{2}$ \\ \vspace{.7cm}
$^{1}$ $^{2}$MINES ParisTech - PLS Research University, Centre de G\'eosciences, Fontainebleau, France\\
‫‬$^{1}$ milad.farshad@mines-paristech.fr\\
$^{2}$ herve.chauris@mines-paristech.fr }

\lefthead{Farshad and Chauris}
\righthead{A PREPRINT, JANUARY 2020}
\maketitle

\begin{abstract}
{\color{black} For quantitative seismic imaging, iterative least-squares reverse time migration is the recommended approach. The existence of an inverse of the forward modelling operator would considerably reduce the number of required iterations. In the context of the extended model, such a pseudo-inverse exists,} built as a weighted version of the adjoint and accounts for the deconvolution, geometrical spreading and uneven illumination. The application of the {\color{black} pseudo-inverse} Born modelling is based on constant density acoustic media, which is a limiting factor for practical applications. To consider density perturbation, {\color{black} we propose and investigate two approaches. The first one is a generalization of a recent study} proposing to recover acoustic perturbations from angle-dependent response of the {\color{black} pseudo-inverse} Born modelling operator. {\color{black} The new version is} based on weighted least-squares objective function. The method not only provides more robust results, but also offers the flexibility to include constrains in the objective function in order to reduce the parameters cross-talk. We also propose an alternative approach based on Taylor expansion that does not require any Radon transform. Numerical examples based on simple and the Marmousi2 models using correct and incorrect background models for the variable density {\color{black} pseudo-inverse} Born modelling, verify the effectiveness of the weighted least-squares method when compared with the other two approaches. The Taylor expansion approach appears to contain too many artifacts for a successful applicability.
\end{abstract}
\graphicspath{{./figures/}}

\section{Introduction}
Seismic migration is a technique for imaging the subsurface structures from observed seismic data. Among different migration algorithms, reverse time migration (RTM) has become the method of choice for seismic imaging in complex geologic structures \citep{rtm1983}. The imaging principle at the core of the RTM algorithms is that the reflectors exist where the upgoing and downgoing wavefields coincide in time and space \citep{Claerbout1971}. {\color{black} We discuss here quantitative seismic imaging algorithms that consider possible density variations.} Conventionally, RTM can be formulated as the adjoint of the linearized forward modelling operator, i.e., adjoint Born modelling operator. {\color{black} It} can only correctly calculate kinematics (phase), and does not preserve amplitudes. In practice, the seismic data are also subject to significant aliasing, noise, irregular source and receiver sampling and finite recording aperture: the final migrated images suffer from artifacts and low-resolution \citep{rtm_comparison}. 

The least-squares migration (LSM) has been proposed to reduce the artifacts mentioned above by reformulating the migration as a least-squares linear inverse problem \citep{Lebras_LSM,Nmeth_LMS}. It historically started with the asymptotic approximation (ray-based formulation) via generalized Radon transform \citep{Beylkin1985}. Migration/inversion formulas have been proposed either based on ray+Born or ray+Kirchhoff linearized approximations to consider quantitative properties of reflection coefficients \citep{Bleistein1987,GRT1997}. In both approaches, the model was split into a unknown reflector/diffractor component and the known background component where the ray tracing was performed. Then, these have been extended to one-way and two-way wave-equation migration \citep{Zhang_LSM2005}. {\color{black} In parallel, a time/space shift extended imaging condition has been proposed for RTM method to potentially decouple the data fitting in RTM from the choice of the velocity model \citep{sava2006extent,Symes2008}.} RTM is considered as the first iteration of LSM. For a quantitative result, least-squares solutions based on RTM (LSRTM) have been proposed to overcome the shortcomings of RTM and to refine images toward true-reflectivity Earth model in terms of phase and amplitudes \citep{Dai_LSRTM2011,Zeng_LSRTM,Zhang_LSRTM2014,Zhang_LSRTM2015}. Two aspects are currently under development: the estimation of multiparameters beyond the constant density acoustic case, and efficient preconditioner for a fast LSRTM.

With the deployment of multi component seismic data, there has been an increasing interest in extending LSRTM to multiparameter imaging in both acoustic and elastic media to provide a better description of wave propagation. Typically, density and attenuation are additional parameters  \citep{Dutta2014Attenuation,Yang2016,sun2018ELSRTM}.

The main drawback of LSRTM is that it should be solved iteratively. Since each iteration requires the application of modelling (demigration) and adjoint (migration) operators, the computational expense of LSRTM can be considerable. Several strategies such as multi source approach with random or linear-phase encoding \citep{Dai_LSRTM2012,Xue_LSRTM2016} have been proposed to increase the efficiency of LSRTM. 

In the context of subsurface offset, recently, different explicit {\color{black} pseudo-inverse} expressions for the Kirchhoff modelling operator \citep{tenkroode} and for the Born modelling operators \citep{symes2015, symes2017a, chauris2017} has been proposed. The {\color{black} pseudo-inverse} operator is an alternative to the adjoint operator and provides quantitative properties within a single iteration. Although the derivations are performed under the high-frequency approximation (ray-theory), the final formulas does not contain any ray quantities but only time and spatial derivatives.  It appears that the technique is very similar to the standard migration scheme, with only additional weights in the imaging operator \citep{symes2015, symes2017a,chauris2017}. These new operators are explicit and simple in terms of implementation.  They can also be used as pre-conditioners (a way to speed up the resolution of the inverse problem) for example for the full waveform inversion and migration velocity analysis \citep{chauris2017,symes2017b,Yubing2018}. The theory of {\color{black} pseudo-inverse} Born modelling is established based on the constant-density acoustic wave equation. However, in reality the density of the Earth is not homogeneous, {\color{black} inhomogeneous or heterogeneous.} Moreover, the amplitude of reflected seismic waves is mostly affected by acoustic impedance contrasts, the product of velocity by density. Therefore, if the P-wave velocity is the only variable parameter in the {\color{black} pseudo-inverse} Born modelling operator while density varies in reality, the inverted parameter will not be correctly estimated.

Very recently, \citet{Dafni2018} extended the method proposed by \citet{symes2017a} to variable density acoustic media. They invert parameters from angle-dependent response of the {\color{black} pseudo-inverse} Born modelling operator based on two traces in the angle-domain. We refer to their method as the ``two-trace'' approach in this paper. {\color{black} Originally, this idea was proposed by \citet{Zhang_MultiRTM2014} in the context of marine acquisition. They delineate the impedance and velocity perturbations from near-angle and far-angle traces of angle-dependent response of amplitude-preserving RTM, respectively \citep{Zhang_MultiRTM2014}.}  Compared to constant density acoustic, variable density inverse RTM provides a better description of the wave propagation and generates more accurate images. Moreover, the resulting models can further be used to study AVO effects that play an important role in lithology analysis and fluid discrimination.

In this paper, we propose a generalization of the method proposed by \citet{Dafni2018} (least-squares method) as well as an alternative approach (Taylor expansion). We show that the final results obtained via \citet{Dafni2018} method slightly depends on the choice of angles for inversion. The least-squares method is based on the use of all traces in the angle-domain, while the Taylor approach is based on the Taylor expansion of the Radon transform around zero-angle. We discuss this method with respect to the dependency on the maximum surface offset and on artifacts in image domain. For all approaches, the starting point is {\color{black} pseudo-inverse} modelling in constant density acoustic media. Here, we extend the \citet{chauris2017} method to variable density acoustic media. We parameterize the subsurface via inverse of the bulk modulus and density and invert these parameters from the angle-dependent response of the {\color{black} pseudo-inverse} Born modelling operator. We compare the three approaches by evaluating the data misfit and inverted parameters reconstruction. We also analyze the sensitivity of the methods to incorrect background models, which is not addressed in \citet{Dafni2018}. In all these approaches, the key factor determining the perfect match in image and shot domains is the pseudo-inverse modelling of constant density acoustic media; their extension is easy to variable density acoustic media.
 
The paper is organized as follows: we first review the preliminaries required for the extension to variable density, namely the variable density acoustic Born modelling, the adjoint and the pseudo-inverse constant density Born modelling, as well as the Radon transform. Then, we explain how to extend the constant density {\color{black} pseudo-inverse} Born operator to variable density acoustic media. We present synthetic examples to compare and discuss the three approaches. Then, we apply the preferred method on the variable density Marmousi2 model. Finally, we discuss the prospects for further development of the {\color{black} pseudo-inverse} Born modelling. {\color{black} This work should be understood as a new step for the applicability of LSRTM, beyond the constant density approximation. By applicability, we mean the derivation of proper pre-conditioners and the possibility in the future to get quantitative LSRTM results within a few iterations only.}

\section{PRELIMINARIES}
We give here a brief review of the concepts and formulas of the variable density acoustic wave equation, the adjoint and the pseudo-inverse operators for the Born modelling, and the Radon transform.
\subsection*{Variable density acoustic Born modelling}
The acoustic Earth model is parameterized with two parameters (at each point), involving p-wave velocity, $V_p$, and density, $\rho$, or their combinations, for instance, p-wave impedance, $I_p = \rho V_p$, or inverse of bulk modulus, $\beta = \frac{1}{\rho V_p^2}$. Each model parameter, for example $\mathbf{m}$, can be considered as the sum of the background model $\mathbf{m_0}$, controlling the kinematics of the wave propagation, and the model perturbation $\delta \mathbf{m}$, creating new types of waves and reflections, where both depends on the spatial coordinates $\mathbf{x} = (x,z)$ {\color{black}\citep{Symes2008}}. By definition, 
\begin{eqnarray} \label{Born}
\mathbf{m}(\mathbf{x}) = \mathbf{m}_0(\mathbf{x}) + \delta \mathbf{m}(\mathbf{x}),
\end{eqnarray}
and, under the Born approximation, we suppose that $\delta \mathbf{m}(\mathbf{x}) \ll \mathbf{m}_0(\mathbf{x})$. The definition of the perturbation model ($\delta \mathbf{m}$) can be extended to depend on more degrees of freedom. The most recent conventional choice for the extension is the subsurface offset ($\mathbf{h}$), introduced as an offset between the sunken source and sunken receiver by \citet{Claerbout1985}. Here, we only consider horizontal subsurface extension as $\mathbf{h} = (h,0)$ for the two-dimensional case. By using this approach, the dimension of the model space becomes the same as the data space (Table \ref{extended}) allowing to compensate for errors in the background model \citep{sava2006extent,Symes2008}. The mathematical expression between physical $\mathbf{m} (\mathbf{x}) $ and extended domains $\mathbf{m} (\mathbf{x},\mathbf{h})$ can be simply defined as
\begin{eqnarray} \label{ext2phy}
\mathbf{m} (\mathbf{x},\mathbf{h}) &=& \mathbf{m} (\mathbf{x}) \delta (\mathbf{h}), \nonumber \\
\mathbf{m}(\mathbf{x}) &=& \int \mathrm{d}\mathbf{h}\, \mathbf{m} (\mathbf{x},\mathbf{h}).
\end{eqnarray}
{\color{black} Equation~\ref{ext2phy} is indeed correct when the background model is correct. For well-focused noise-free data, extracting the image at $\mathbf{h}=0$ could be reasonably accurate. However, it is more robust to sum over $\mathbf{h}$-axis. In other word, one might expect the estimated physical image so obtained to be less sensitive to incoherent or numerical noise \citep{symes2015}.}
\begin{table}[h]\caption{Dimension of the data and model domains. $s$ and $r$ are the source and receiver coordinates; $t$ is the time; $x$ , $y$ and $z$ are the spatial coordinates; and $h$ is the subsurface offset.}

\centering
\renewcommand{\arraystretch}{1.2}
\begin{tabular}{cccc}
\Cline{2pt}{1-4}
Dimension & Data domain           & Physical model domain & Extended model domain   \\
\hline
1D        & $t$                   & $z$                   & $z$                   \\
2D        & $(s,r,t)$             & $(x,z)$               & $(x,z,h)$              \\
3D        & $(s_x,s_y,r_x,r_y,t)$ & $(x,y,z)$             & $(x,y,z,h_x,h_y)$    \\
\hline
\end{tabular}\label{extended}
\end{table}
We also denote the source and receiver positions as $\mathbf{s}$ and $\mathbf{r}$, respectively. {\color{black} We parameterize the subsurface via inverse of the bulk modulus and density.} The solution of the 2-D acoustic scattered wavefield under the Born approximation can be expressed by introducing the reference Green's function $G_0$ in a given background model $\mathbf{m}_0$. An integral operator expression for the extended Born modelling operator $\mathcal{B}$ \citep{Symes2008} can be written as an integral over all scatter positions: 

\begin{eqnarray} \label{multi_v_rho}
d(\mathbf{s},\mathbf{r},\omega) = \mathcal{B}[\mathbf{m_0}]\delta\mathbf{m}(\mathbf{s},\mathbf{r},\omega) &=& \Omega(\omega) \int \mathrm{d}\mathbf{x}\, \mathbf{F}_{(\beta,\rho)}(\mathbf{x},\mathbf{h},\omega ;\mathbf{s},\mathbf{r}) \delta\mathbf{m}_{(\beta,\rho)}(\mathbf{x},\mathbf{h}),
\end{eqnarray}
where $\omega$ is the angular frequency, $\Omega(\omega)$ is the source spectrum, $\mathbf{F}_{(\beta,\rho)}$ is the {\color{black} standard} modelling vector given as {\color{black}\citep{Symes2008}:}

\begin{eqnarray} \label{F}
\mathbf{F}_{(\beta,\rho)}(\mathbf{x},\mathbf{h},\omega ;\mathbf{s},\mathbf{r})=
\begin{bmatrix}[2]
-(i\omega)^2 G_0(\mathbf{s},\mathbf{x} - \mathbf{h},\omega)G_0(\mathbf{x} + \mathbf{h},\mathbf{r},\omega) \\
 \frac{1}{\rho_0^2} \nabla G_0(\mathbf{s},\mathbf{x} - \mathbf{h},\omega)\cdot \nabla G_0(\mathbf{x} + \mathbf{h},\mathbf{r},\omega)
\end{bmatrix}^T,
\end{eqnarray}
$\delta\mathbf{m}_{(\beta,\rho)}$ is the model vector given as:
\begin{eqnarray}
\delta\mathbf{m}_{(\beta,\rho)}(\mathbf{x},\mathbf{h})=
\begin{bmatrix}[2]
\delta \beta(\mathbf{x},\mathbf{h}) \\
\delta \rho(\mathbf{x},\mathbf{h})
\end{bmatrix},
\end{eqnarray}
and $T$ denotes the transpose operator. The composition $\mathcal{B}[\mathbf{m_0}]\delta\mathbf{m}$ denotes computed data $d$ in $(\mathbf{m_0},\delta\mathbf{m})$. Since different parameter classes have different physical units and nature, they can have different influence on the data. Note that the influence of one parameter on the data depends on the other parameters involved in the subsurface parameterization. Generally, this is referred to as parameter cross-talk, meaning that parameters are more or less coupled \citep{virieuxfwi}. The diffraction or radiation pattern can give some insight into influence of the parameterization in the data as a function of the diffraction angle \citep{Operto2013parametrization}. The analytical expression for the diffraction pattern can be derived in the framework of the Ray+Born approximation \citep{Forgues}. Thus, under the high-frequency approximation, the modelling vector $\mathbf{F}_{(\beta,\rho)}$ (Equation~\ref{F}) can be rewritten as :
\begin{eqnarray} \label{F2}
\mathbf{F}_{(\beta,\rho)}(\mathbf{x},\mathbf{h},\omega ;\mathbf{s},\mathbf{r}) &=&
\begin{bmatrix}[2]
-(i\omega)^2 G_0(\mathbf{s},\mathbf{x} - \mathbf{h},\omega)G_0(\mathbf{x} + \mathbf{h},\mathbf{r},\omega)  \\
\frac{\beta_0(i\omega)^2 }{\rho_0}G_0(\mathbf{s},\mathbf{x} - \mathbf{h},\omega)G_0(\mathbf{x} + \mathbf{h},\mathbf{r},\omega)\cos(2\gamma)
\end{bmatrix}^T,\nonumber \\
&=& (i\omega)^2 G_0(\mathbf{s},\mathbf{x} - \mathbf{h},\omega)G_0(\mathbf{x} + \mathbf{h},\mathbf{r},\omega)
\begin{bmatrix}
-1 &  \frac{\beta_0}{\rho_0} \cos(2\gamma)
\end{bmatrix},
\end{eqnarray}
where $\gamma$ denotes the diffraction angle. Figure~\ref{fig:radiation} shows the graphic representation of diffraction pattern for $\mathbf{m}_{(\beta,\rho)}$ as a function of the diffraction angle. As expected from Equation~\ref{F2}, the amplitude of scattered wavefield by $\beta$ perturbation in the ${(\beta,\rho)}$ parameterization shows an isotropic pattern (red line in Figure~\ref{fig:radiation}), while the amplitude by $\rho$ perturbation shows an increasingly decreasing pattern from the small $\gamma$ to the wider ones (blue line in Figure~\ref{fig:radiation}).
\plot{radiation}{width=.4\columnwidth}
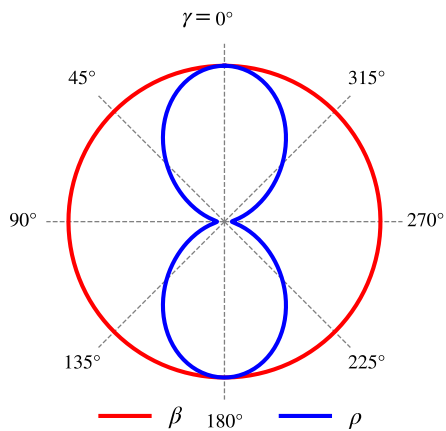
{The analytical diffraction pattern for an acoustic medium parameterized by $(\beta,\rho)$. }

Equation~\ref{F2} provides a linear relationship between the data and the $(\delta\beta,\delta\rho)$ perturbations. In the next section, we review the {\color{black} pseudo-inverse} operator providing optimal $\delta\beta(\mathbf{x},\mathbf{h})$ from observed data in the case when $\delta\rho=0$. Then we introduce the Radon transform to handle angle $\gamma$ appearing in Equation~\ref{F2} when $\delta\rho \neq 0$. 
\subsection*{Pseudo-inverse Born modelling}
RTM is one of the most powerful seismic imaging methods in complicated areas. The classical form of RTM operator can be expressed as the correlation between the forward and back-propagation of source and receiver wavefields, respectively \citep{migbednar}. By considering constant density acoustic media ($\delta \rho = 0$), for a specific background model ($\beta_0$), migration is introduced by minimizing the misfit between the observed $(d^{obs})$ and computed data ($d(\xi)$) at each shot position as:

\begin{eqnarray} \label{initial_cost}
J_0(\xi) = \frac{1}{2} \norm{d(\xi) - d^{obs}}^2,
\end{eqnarray}
where $\xi=\delta \beta$ is the extended reflectivity and {\color{black}$\norm{\,.\,}^p$ denotes the $\ell_p$-norm.}
The operator of the migration for determining $\xi$ can be defined by deriving the gradient of the Equation~\ref{initial_cost} with respect to $\xi$ as $(\partial J_0 / \partial \xi) |_{\xi=0}$ \citep{Lailly1983,Tarantola1984}, yielding
\begin{eqnarray}\label{adjborn}
\xi = \mathcal{B}^{T}(d^{obs})(\mathbf{x},\mathbf{h}) =  -\int \mathrm{d}\mathbf{s}\,\mathrm{d}\mathbf{r}\,\mathrm{d}\omega \,(i\omega)^2 \Omega^{*} (\omega) G_0^*(\mathbf{s},\mathbf{x} - \mathbf{h},\omega) d^{obs}(\mathbf{s},\mathbf{r},\omega)G_0^*(\mathbf{x} + \mathbf{h},\mathbf{r},\omega),
\end{eqnarray}
where ${}^*$ denotes the complex conjugate. RTM (Equation~\ref{adjborn}) is the adjoint of the Born modelling operator ($\mathcal{B}^T$), whereas inversion ($\mathcal{B}^\dagger$) is its asymptotic inverse. Here, for {\color{black} pseudo-inverse} Born modelling, we considered the method proposed by \citet{chauris2017} which is close to the one presented in \citet{symes2017a}. Both derivations determine an {\color{black} pseudo-inverse} $\mathcal{B}^\dagger$ of the extended Born modelling operator $\mathcal{B}$ such that $\mathcal{B}^\dagger \mathcal{B} \approx I$, where $I$ is the identity operator. However, \citet{symes2017a} directly apply the stationary phase approximation on $\mathcal{B}^\dagger \mathcal{B}$, whereas \citet{chauris2017} use a linearization of the phase of $\mathcal{B}^\dagger \mathcal{B}$, leading to different final formulations for $\mathcal{B}^\dagger$ even if both formulations are valid in an asymptotic sense \citep{chauris2018EAGE}. Following the work by \citet{chauris2017}, the pseudo-inverse formula for the extended Born modelling operator in a constant density acoustic media can be written as:
\begin{eqnarray}\label{iborn}
\xi = \mathcal{B}^{\dagger}(d^{obs})(\mathbf{x},\mathbf{h}) \simeq 32\frac{\beta_0}{\rho_0^3}\partial_{z} \int \mathrm{d}\mathbf{s}\,\mathrm{d}\mathbf{r}\,\mathrm{d}\omega \,\frac{ \Omega^{\dagger} (\omega)}{(i\omega)} \partial_{s_{z}}G_0^*(\mathbf{s},\mathbf{x} - \mathbf{h},\omega) d^{obs}(\mathbf{s},\mathbf{r},\omega)\partial_{r_{z}}G_0^*(\mathbf{x} + \mathbf{h},\mathbf{r},\omega),
\end{eqnarray}
where $\Omega^{\dagger} = \frac{\Omega^*}{\norm{\Omega}^2}$ is the inverse of the seismic source. In terms of implementation, Equation~\ref{iborn} is close to the standard migration algorithm (Equation~\ref{adjborn}), with three main modifications: (1) applying vertical derivative with respect to source and receiver positions to the Green's functions; (2) using the inverse version of the source wavelet instead of adjoint version; (3) applying a first-order integration in time before cross-correlation instead of a second-order derivative and finally a vertical derivative to the result of the cross-correlation. These modifications account to applying the following weights, respectively: (1) cosines of take-off angles at the sources and receivers positions; (2) deconvolution of the source wavelet; (3) cosines of the half-opening angle at the image point. A larger weight is given to small scattering angles and short surface offsets \citep{chauris2017}. In constant density acoustic media, if the investigated background model is correct, the energy focuses around the zero-subsurface offset. Otherwise, the extended domain allows to compensate for errors in the background velocity model by defocusing the energy in the extended reflectivity. Although, there would be no physical sense of real perturbation model, it is still possible to reconstruct the observed data from the extended reflectivity \citep{Symes2008}. For example, Figure~\ref{fig:cig_zero_rho}a shows the observed data for a single layer Earth model in a constant density acoustic media. The corresponding common-image gather (CIG) for correct ($v = 2400$ m/s) and incorrect ($v = 2700$ m/s)  background models are also shown in Figure~\ref{fig:cig_zero_rho}b-c, respectively. It is worth to note that the high (low) velocity results in upward (downward) curvature in CIG domain, as it is in Figure~\ref{fig:cig_zero_rho}c. To evaluate the effect of the extended domain, we first reconstruct the shot for the correct and incorrect background models in the extended domain (Figure~\ref{fig:shots_zero_rho}a-b). The phase and amplitude are correctly retrieved for both cases. Then, we reconstruct the data for incorrect background model in the physical domain by simply summing CIG over all $\mathbf{h}$ values  (see Equation~\ref{ext2phy}, Figure~\ref{fig:shots_zero_rho}c). Careful examination on the extracted traces indicates that the phase is not well retrieved (Figure~\ref{fig:shots_zero_rho}c).

\plot{cig_zero_rho}{width=.75\columnwidth}
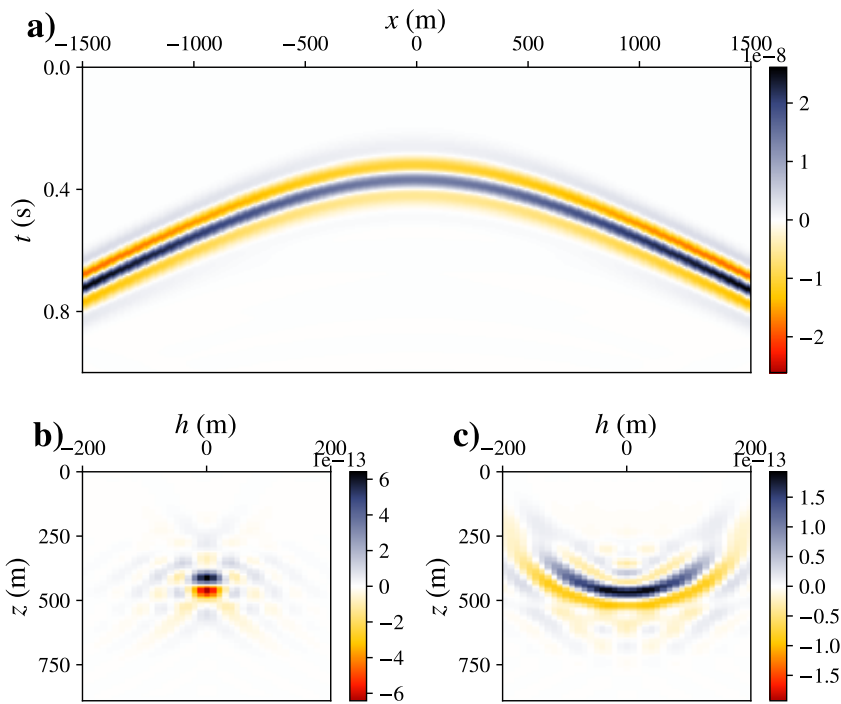
{Constant density acoustic media. a) Observed data for a single layer Earth model and the corresponding CIG inverted via {\color{black} pseudo-inverse} Born modelling operator (Equation~\ref{iborn}) with b) correct and c) incorrect background model (high velocity).}

\plot{shots_zero_rho}{width=.75\columnwidth}
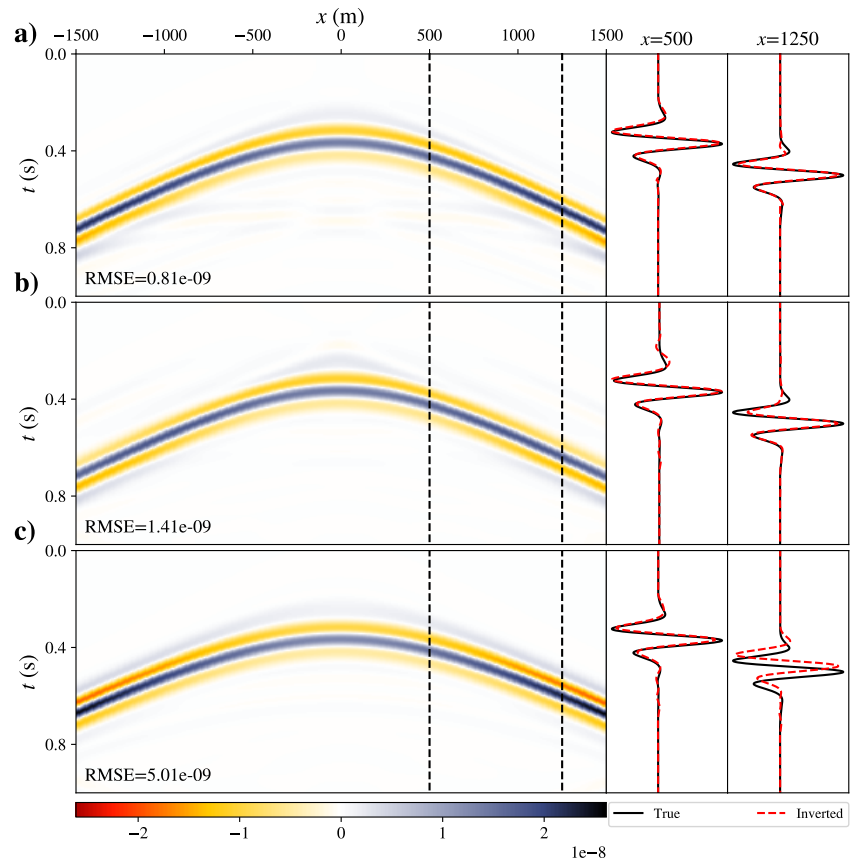
{Left: reconstructed shots with a) correct velocity extended domain, b) incorrect velocity extended domain and c) incorrect velocity physical domain. The RMS error between synthetic and observed data is written on each panel. right: extracted traces for near and far offsets.}

In variable density acoustic media, the energy of the incident wave is partitioned at each boundary based on the contrasts in velocity and density properties across the boundary. The important aspect to note is that reflection amplitude for each boundary depends on the angle of incidence. This phenomenon is well known as amplitude variation with angle (AVA) \citep{avo2014}. An example of observed data for a single layer Earth model in a variable density acoustic media and its reflection coefficient as a function of angle for the interface is shown Figure~\ref{fig:cig_variable_rho}a-b, respectively. The velocity and density pairs for above and below the interface are ($2500$ m/s, $1950$ gr/dm$^3$) and ($2320$ m/s, $2200$ gr/dm$^3$), respectively. In such media, using Equation~\ref{iborn} for inverting reflectivity will defocus the energy in extended domain to compensate the effect of AVA, even with correct background model, as can be seen in Figure~\ref{fig:cig_variable_rho}c. Although it may still reconstruct the observed data, there would be no physical sense of velocity or density perturbation since the energy is defocused (Figure~\ref{fig:cig_variable_rho}c.). 

As we showed in Equation~\ref{F2}, the combination of the ray theory and the Born approximation provides quantitative estimations of the multiparameter inversion. It means that the inverted $\xi(\mathbf{x},\mathbf{h})$ in the extended domain (Equation~\ref{iborn}), can be decomposed into two physical parameters $(\delta \beta(\mathbf{x}),\delta \rho(\mathbf{x}))$ based on the diffraction pattern of the specific parameterization. The equation for this relation can be written as:
\plot{cig_variable_rho}{width=.75\columnwidth}
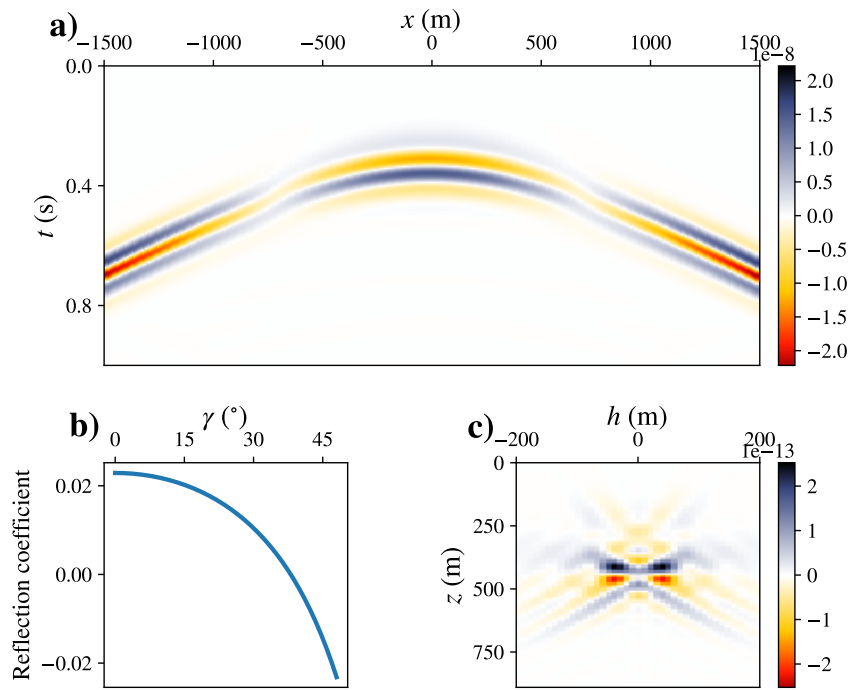
{Variable density acoustic media. a) Observed data for a single layer Earth model and the corresponding b) AVA for the interface, and c) CIG inverted via {\color{black} pseudo-inverse} Born modelling operator (\ref{iborn}) with correct background model.}

\begin{eqnarray} \label{xi1}
\xi_\beta (\mathbf{x},\mathbf{h}) \cong \delta \beta(\mathbf{x}) \delta (\mathbf{h}) - \frac{\beta_0}{\rho_0}\cos(2\gamma) \delta \rho(\mathbf{x}) \delta (\mathbf{h}), 
\end{eqnarray}
where $\delta()$ is the Dirac delta function. In the case $\delta\rho(\mathbf{x}) = 0$, the $\delta \beta(\mathbf{x})$ can be reconstructed from $\xi_\beta (\mathbf{x},\mathbf{h})$ by simply summing over all $\mathbf{h}$ values ($\delta \beta(\mathbf{x})=\int \mathrm{d}\mathbf{h}\,\xi_\beta (\mathbf{x},\mathbf{h})$). For variable density, based on the fact that inversion of $\delta \rho(\mathbf{x})$ in Equation~\ref{xi1} is angle-dependent, a transformation of subsurface offset to scattering angle is required \citep{Dafni2018}. In the next section, we review how to transform the extended offset domain CIG to angle domain CIG via the Radon transform.
\subsection*{Angle domain CIGs}
Angle gathers are the main ingredient of the AVO/AVA analysis which  can give us reliable estimates of the Earth parameters, such as, P-wave velocity ($V_p$), S-wave velocity ($V_s$), density ($\rho$), or different combinations of them. These parameters can further be used to provide information about reservoir parameters, namely, lithology, porosity and fluid content \citep{Castagna1994,avo2014}. Angle gathers can be obtained using wave-equation techniques either based on wavefield methods \citep{debruin,sava2003RT,Biondi2004,Sava2011,Sava2013,Dafni2016} or ray methods \citep{Brandsberg}. The fact that the wavefield methods can accurately image complex geologic structures comparing to ray methods, makes wavefield methods superior to ray methods. The wavefield methods can be applied either for shot-profile migration, shot-geophone migration, or prestack images after migration. Since in the first two method the angle gathers are evaluated from wavefield prior to imaging, they referred as ``data-space methods''. The angle gathers obtained from these methods are a function of offset ray parameter. \citet{sava2003RT} showed that angle gathers can also be obtained from migrated images with a process which is completely detached from migration. The main advantages of their method are that the angle gathers are produced as a function of the reflection angle, which is not the case in data-space methods, and also with much less computational cost compared to data-space methods. This process is based on performing the Radon transform (slant stack integral) on extended CIGs.  The Radon transform formula can be written in 2D as:
\begin{eqnarray} \label{RT}
\mathcal{R}_{\xi}(x,z,\gamma) = \int \mathrm{d}h\,\xi(x,z+h\tan\gamma,h),
\end{eqnarray}
\\
where $\tan\gamma$ is the trajectory of integration. Figure~\ref{fig:RT} illustrates the integration path for different point positions in a CIG panel. {\color{black} Note that it is commonly assumed that the RTM and the ray-based angle ($\gamma$ in Equation~\ref{F2} and \ref{RT}) are the same. \citet{Montel2011angle} showed that this is not necessarily the case for incorrect background models.}

\plot{RT}{width=1\columnwidth}
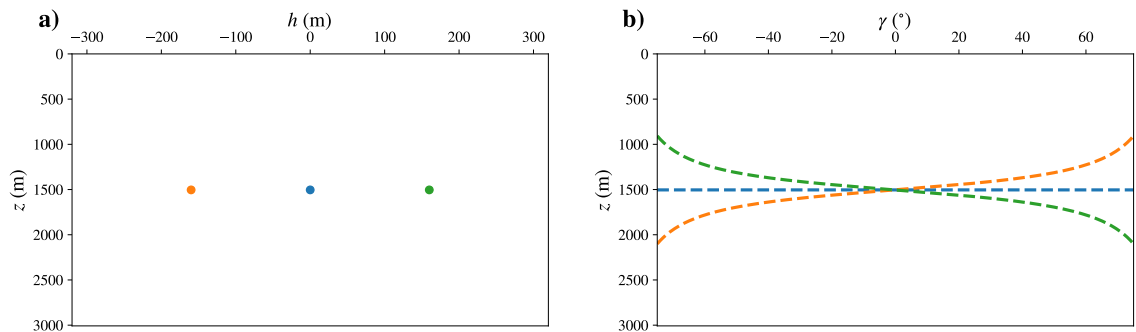
{a) a CIG with three nonzero points and b) its Radon transform showing the path of integration for different points.}

We now have the ingredients to take into account variable density, i.e. to determine $(\delta\beta(\mathbf{x}),\delta\rho(\mathbf{x}))$ from $\xi_{\beta}(\mathbf{x},\mathbf{h})$ (Equation~\ref{xi1}). As the inversion formula is derived in data domain (\ref{initial_cost}), we do not expect a unique solution, but the reconstructed data from $(\delta\beta(\mathbf{x}),\delta\rho(\mathbf{x}))$ should match the observed data. {\color{black} In other word, we may have different combinations of $\delta\beta(\mathbf{x})$ and $\delta\rho(\mathbf{x})$, leading to approximately the same data fit.} Note that $\xi_{\beta}(\mathbf{x},\mathbf{h})$ is defined in the extended domain, whereas $(\delta\beta(\mathbf{x}),\delta\rho(\mathbf{x}))$ are defined in the physical domain.
\section{Variable density pseudo-inverse modelling}
Here, we first review the two-trace method proposed by \citet{Dafni2018}. Then, we derive two new schemes for variable density {\color{black} pseudo-inverse} Born modelling. The first one is based on the generalization of the two-trace method, referred to as weighted least-squares (WLS), and the second one is based on the Taylor expansion of the Radon transform around $\gamma=0$. {\color{black} As detailed below, the latter does not require any forward/inverse Radon transform, with a straightforward implementation. We explain its derivation and discuss its applicability.}
\subsection{Two-trace method}
Recently, multiparameter inversion method for bulk modulus and density perturbation was proposed by \citet{Dafni2018} based on the analysis of two traces within the Radon domain. They used a similar formula as Equation~\ref{xi1}, to invert the perturbations from the angle-dependent response of the {\color{black} pseudo-inverse}. By applying the Radon transform on $\xi$ (Equation~\ref{xi1}):
\begin{eqnarray} \label{WLS}
{\mathcal{R}}_{\xi}(\mathbf{x},\gamma) = \delta\beta(\mathbf{x}) - \frac{\beta_0}{\rho_0}\cos(2\gamma)\delta\rho(\mathbf{x}),
\end{eqnarray}
they propose to choose two traces in angle-domain, as:
\begin{eqnarray} \label{2trace}
a(\mathbf{x}) &=& \mathcal{R}_{\xi}(\mathbf{x},0) = \delta\beta(\mathbf{x}) - \frac{\beta_0}{\rho_0}\delta\rho(\mathbf{x}), \\
b(\mathbf{x}) &=& \mathcal{R}_{\xi}(\mathbf{x},\gamma') = \delta\beta(\mathbf{x}) - \frac{\beta_0}{\rho_0}\cos(2\gamma')\delta\rho(\mathbf{x}),
\end{eqnarray}
and then calculate $\delta \beta$ and $\delta \rho$ by a system of two linear equations as:
\begin{eqnarray}\label{2tracesystem}
\delta \rho (\mathbf{x}) &=& \frac{\rho_0}{\beta_0} \frac{b(\mathbf{x}) - a(\mathbf{x})}{1 - \cos(2\gamma')},\nonumber \\
\delta \beta (\mathbf{x}) &=& \frac{b(\mathbf{x}) - \cos(2\gamma')a(\mathbf{x})}{1 - \cos(2\gamma')}.
\end{eqnarray}
It should be noted that, inversion results for this method is angle dependent, meaning that the results might slightly change by choosing different value for $\gamma'$.

\subsection{Weighted least-squares (WLS) method}
In this paper, in order to get more robust results, we generalize the approach to consider all traces in the Radon domain. Thus, we consider the whole AVA response rather than simple expansions in terms of two traces. The objective function {\color{black} for any given $\mathbf{x}=(x,z)$} for optimal $\delta\beta(\mathbf{x})$ and $\delta \rho(\mathbf{x})$ can be defined as the least-squares differences between the computed and observed $\mathcal{R}_{\xi}$ over the all angles as:
\begin{eqnarray} \label{J3}
J_{\delta\beta,\delta\rho} = \frac{1}{2} \norm{(\delta\beta(\mathbf{x}) - \frac{\beta_0}{\rho_0}\cos(2\gamma)\delta\rho(\mathbf{x})) -\mathcal{R}_{\xi}(\mathbf{x},\gamma)}_{W}^2,
\end{eqnarray}
where $W(\mathbf{x},\gamma)$ is a weighting mask defined as:
\begin{eqnarray} \label{W}
W(\mathbf{x},\gamma) = \begin{cases} 1 ,& \mbox{if  } |\gamma | \leq \alpha \tan^{-1}(\frac{x_{max}}{z})  \\
0  ,& \mbox{otherwise}
\end{cases}.
\end{eqnarray}
We define this mask based on the acquisition geometry to remove the artifacts in the angle-domain at large angles. Thus, for each depth, we only consider the angles that would indeed be recorded. {\color{black} This is based on constant assumption.} In Equation~\ref{W}, $\alpha$ is a value close to 1, which is essential to avoid the artifacts due to the finite sampling of the offset axis. {\color{black} In practice, there is no need to have a very precise definition of $\alpha$. We indeed choose this parameter to guarantee that we mainly include the specular events in the angle-domain CIG.} Figure~\ref{fig:mask} shows the weighting mask for $x_{max} = 2250$ m, $\alpha=0.85$ and $\alpha=1$. Outside of these boundaries, the angle domain only contains the artifacts corresponding to the artifacts in CIG domain at large subsurface offset values. 

In order to derive the optimal $\delta \beta (\mathbf{x})$ and $\delta \rho(\mathbf{x})$, we compute the gradient of the objective function $J$ (Equation~\ref{J3}) with respect to the model parameters $\delta \beta(\mathbf{x})$ and $\delta \rho(\mathbf{x})$ as:

\begin{eqnarray} \label{J_beta}
\frac{\partial J}{\partial \delta\beta} &=&  \int \mathrm{d} \gamma W(\mathbf{x},\gamma) \Big( \delta \beta(\mathbf{x}) - \frac{\beta_0}{\rho_0} \cos(2\gamma) \delta\rho(\mathbf{x}) -\mathcal{R}_{\xi}(\mathbf{x},\gamma) \Big) = 0 ,\nonumber\\
\frac{\partial J}{\partial \delta\rho} &=&  -\int \mathrm{d} \gamma W(\mathbf{x},\gamma) \frac{\beta_0}{\rho_0} \cos(2\gamma) \Big( \delta \beta(\mathbf{x}) - \frac{\beta_0}{\rho_0} \cos(2\gamma) \delta\rho(\mathbf{x}) -\mathcal{R}_{\xi}(\mathbf{x},\gamma) \Big) = 0.
\end{eqnarray}
Equation~\ref{J_beta} can hence be written in a matrix formulation as:

\begin{eqnarray} \label{J_mat}
\begin{bmatrix}[2]
\int \mathrm{d} \gamma W(\mathbf{x},\gamma) &  - \int \mathrm{d} \gamma W(\mathbf{x},\gamma) \frac{\beta_0}{\rho_0} \cos(2\gamma) \\
- \int \mathrm{d} \gamma W(\mathbf{x},\gamma) \frac{\beta_0}{\rho_0} \cos(2\gamma) &  \int \mathrm{d} \gamma W(\mathbf{x},\gamma) (\frac{\beta_0}{\rho_0})^2 \cos^2(2\gamma) \\
\end{bmatrix}
\begin{bmatrix}[2]
\delta \beta(\mathbf{x}) \\
\delta \rho(\mathbf{x})
\end{bmatrix} 
=
\begin{bmatrix}[2]
\int \mathrm{d} \gamma W(\mathbf{x},\gamma) \mathcal{R}_{\xi}(\mathbf{x},\gamma)  \\ 
-\int \mathrm{d} \gamma W(\mathbf{x},\gamma) \frac{\beta_0}{\rho_0} \cos(2\gamma) \mathcal{R}_{\xi}(\mathbf{x},\gamma) 
\end{bmatrix}.
\end{eqnarray}
Solution of Equation~\ref{J_mat} leads to the determination of an optimal $\delta \beta(\mathbf{x})$ and $\delta \rho(\mathbf{x})$ in physical domain.

\plot{mask}{width=.6\columnwidth}
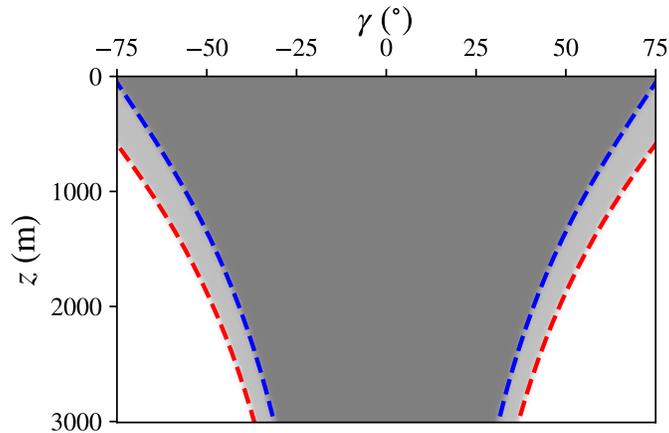
{Weighting mask $W(\mathbf{x},\gamma)$. The red and blue dashed lines defines the $W$ for $\alpha=1$ and $\alpha=0.85$, respectively. }

\subsection{Taylor expansion}
Here, we propose an alternative approach based on Taylor expansion along the angle $\gamma$ of the Radon transform around $\gamma=0$. The Taylor series provides a sum of the terms which will approximate the function. Thus, the Taylor expansion of the Radon transform (Equation~\ref{RT}) evaluated around $\gamma=0$ can be written as:

\begin{eqnarray} \label{taylorRt}
\mathcal{R}_{\xi}(\mathbf{x},\gamma) \simeq \mathcal{R}_{\xi}(\mathbf{x},0) + \gamma \frac{\partial \mathcal{R}_{\xi}}{\partial \gamma}\bigg|_{\gamma=0} + \frac{1}{2}\gamma^2 \frac{\partial^2 \mathcal{R}_{\xi}}{\partial \gamma^2}\bigg|_{\gamma=0}, 
\end{eqnarray}
where the first term is:
\begin{eqnarray} \label{taylorRt_t1}
\mathcal{R}_{\xi}(\mathbf{x},0) =  \int \mathrm{d}h\,\xi(x,z,h),
\end{eqnarray}
 the second term is:
\begin{eqnarray} \label{taylorRt_t2}
\frac{\partial \mathcal{R}_{\xi}}{\partial \gamma}\bigg|_{\gamma=0} = \int \mathrm{d}h\,h \frac{\partial \xi}{\partial z}(x,z,h),
\end{eqnarray}
and the third term is:
\begin{eqnarray} \label{taylorRt_t3}
\frac{\partial^2 \mathcal{R}_{\xi}}{\partial \gamma^2}\bigg|_{\gamma=0} = \int \mathrm{d}h\, h^2 \frac{\partial^2 \xi}{\partial z^2}(x,z,h),
\end{eqnarray}
and also the Taylor expansion of the $\cos(2\gamma)$ evaluated around $\gamma=0$ can be written as:
\begin{eqnarray} \label{taylorcos}
\cos(2\gamma) \simeq 1 - 2\gamma^2.
\end{eqnarray}
Note that the second term (Equation~\ref{taylorRt_t2}) consists of integration of vertical derivative of $\xi(x,z,h)$ multiplied with $h$ including negative and positive values. This leads the second term to be negligible comparing to the first and third terms (Equation~\ref{taylorRt_t1} and \ref{taylorRt_t3}, respectively). Substituting Equation~\ref{taylorRt} and \ref{taylorcos} into Equation~\ref{2tracesystem} gives

\begin{eqnarray}\label{taylorsystem}
\delta \rho (\mathbf{x}) &\simeq & \frac{\rho_0}{4\beta_0} \frac{\partial^2 \mathcal{R}_{\xi}}{\partial \gamma^2}\bigg|_{\gamma=0}=\frac{\rho_0}{4\beta_0}\int \mathrm{d}h \,h^2 \frac{\partial^2 \xi}{\partial z^2}(x,z,h), \nonumber \\
\delta \beta (\mathbf{x}) &\simeq & \mathcal{R}_{\xi}(\mathbf{x},0) + \frac{1}{4} \frac{\partial^2 \mathcal{R}_{\xi}}{\partial \gamma^2}\bigg|_{\gamma=0} = \int \mathrm{d}h \Big[\xi(x,z,h) + \frac{h^2}{4}  \frac{\partial^2 \xi}{\partial z^2}(x,z,h)\Big]. 
\end{eqnarray}
The advantage of this formulation is that it does not require any application of the Radon transform. As can be seen, the $\delta\rho(\mathbf{x})$ only depends on second derivation of the $\xi(x,z,h)$ weighted with $h^2$. This makes it sensitive to the artifacts in CIG domain at large subsurface offset values. Moreover, the Taylor expansion of the Radon transform is applied around $\gamma=0$ meaning short-offset acquisition. However, in seismic we prefer long-offset acquisitions to be able to reconstruct the large and intermediate wavelengths of wavefield. 


\newpage
\section{NUMERICAL EXPERIMENTS}
We present several numerical examples from simulated data to compare the performance of the different methods in term of accuracy, i.e., the ability for data reconstruction. For the first example, we used the same model used by \citet{Dafni2018}. We analyze the results for correct and incorrect background models for this example. In the second example, we test the application of proposed method on variable density Marmousi2 dataset.

\subsection*{Simple model}
This model consists of 4 horizontal interfaces, which each of them shows different type of AVA signature. The exact $\beta$ and $\rho$ model is shown in Figure~\ref{fig:model_1}a-b. The reflection coefficient corresponding to each event is also shown in Figure~\ref{fig:model_1}-c. As can be seen, a strong AVA effect is expected for the first and second interfaces. The sign of the amplitude for the first event changes with angle from negative to positive and it vanishes around $\gamma=20^{\circ}$. For the second event, the amplitude starts from zero at $\gamma=0^{\circ}$ and decreases with angle. The model is discretized on a $189 \times 189$ grid with $16$ m spacing in vertical and horizontal direction. The source wavelet for the simulation is a Ricker signal centered at $4.6$ Hz (maximum frequency is $11.5$ Hz). Each shot is recorded on $189$ channels during $3$ s with a $3.5$ ms time interval. {\color{black} By considering correct background models (Figure~\ref{fig:cig_correct_model1}a-b), the inverted $\xi$ and its Radon transform are shown in Figure~\ref{fig:cig_correct_model1}c-d.} As expected, even by considering correct background models and inversion method, the energy in CIG domain is defocused due to AVO/AVA phenomenon (mainly the shallowest event). {\color{black} As illustrated in Figure~\ref{fig:RT}, energy at different lags in offset domain CIGs, leads to different integral paths in angle domain CIGs. This leads to have more amplitude distortion at higher angles in angle domain CIGs.}

\plot{model_1}{width=1\columnwidth}
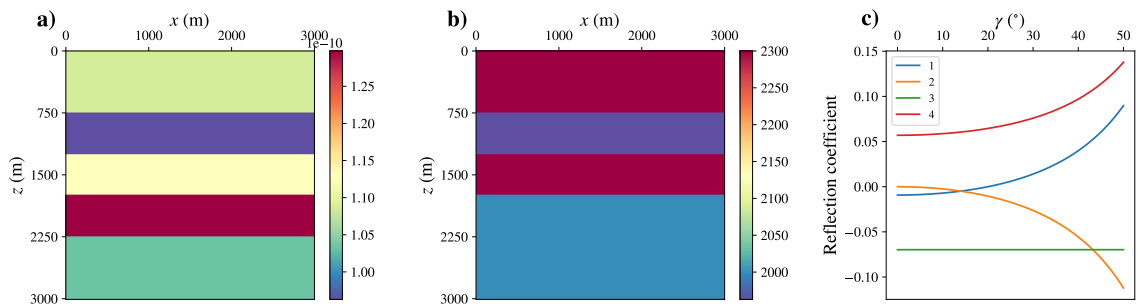
{exact a) $\beta$ (s$^2$dm$^3$/m$^2$gr), b) $\rho$ (gr/dm$^3$) and c) reflection coefficient as a function of angle for each interface corresponding to model 1.}

\plot{cig_correct_model1}{width=1\columnwidth}
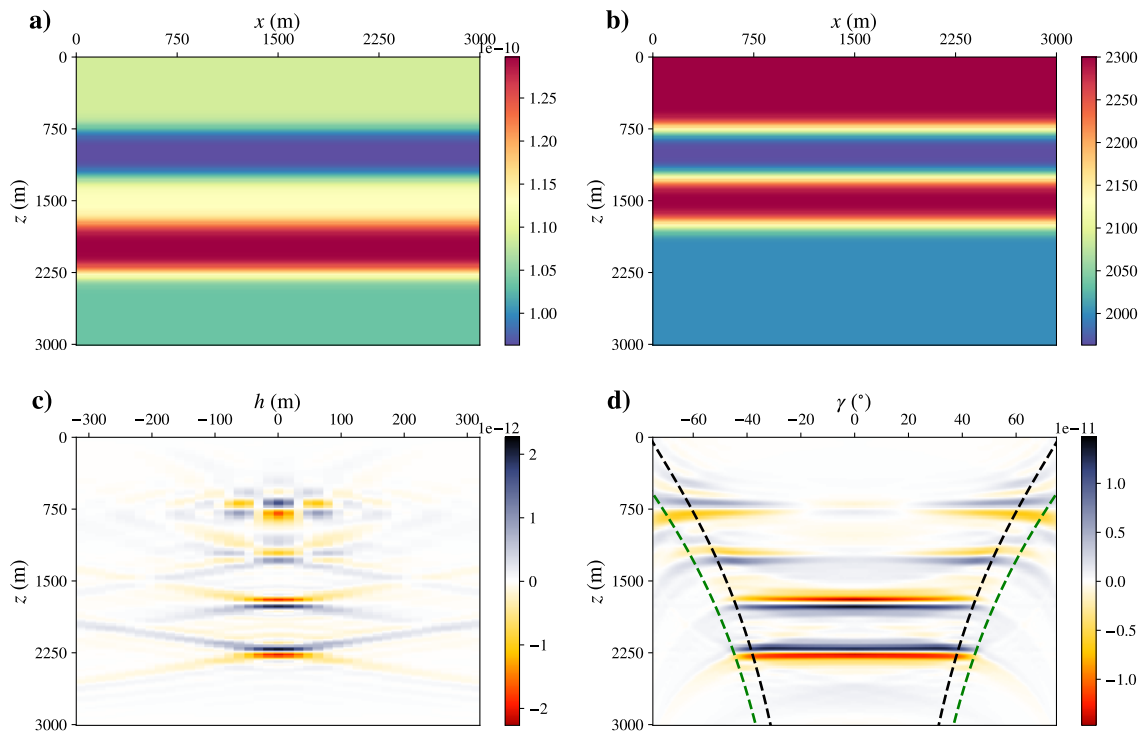
{The initial correct background a) $\beta_0$ and b) $\rho_0$ models used for pseudo-inverse Born modelling. Inverted c) $\xi$ and its d) $\mathcal{R}_{\xi}$ at position $x=1500$ m corresponding to Figure~\ref{fig:model_1}.}

First, we apply the extended Born modelling on the inverted physical models to re-simulate the data. The observed and computed shots, and the extracted traces for near/far offsets for each shot are calculated (Figure~\ref{fig:shots_correct_model1}). {\color{black} The colorbar scale for reconstructed data is the same as the one for the observed data.} There is a perfect match between observed and computed shots both in terms of phase and amplitudes. The Root Mean Squared Error (RMSE) between observed and reconstructed shot via each method is also written on each panel showing the superiority of the WLS method. As can be seen in the extracted traces, there is mismatch for Taylor expansion method specially in shallow part of the data containing larger angle information. This indeed is caused by the approximation of the Radon transform around $\gamma=0$. It is interesting to compare the observed data and the reconstructed data based on short-offset acquisition where the Taylor expansion should be more applicable. Therefore, we apply the modelling on the same model but with considering short-offset acquisition (Figure~\ref{fig:shots_noRT_short}). There is indeed better match in both shallow and deep parts. Note that the RMSE between the observed and reconstructed shot in the short-offset acquisition is 3 times less than the long-offset acquisition for Taylor expansion method.

\plot{shots_correct_model1}{width=.75\columnwidth}
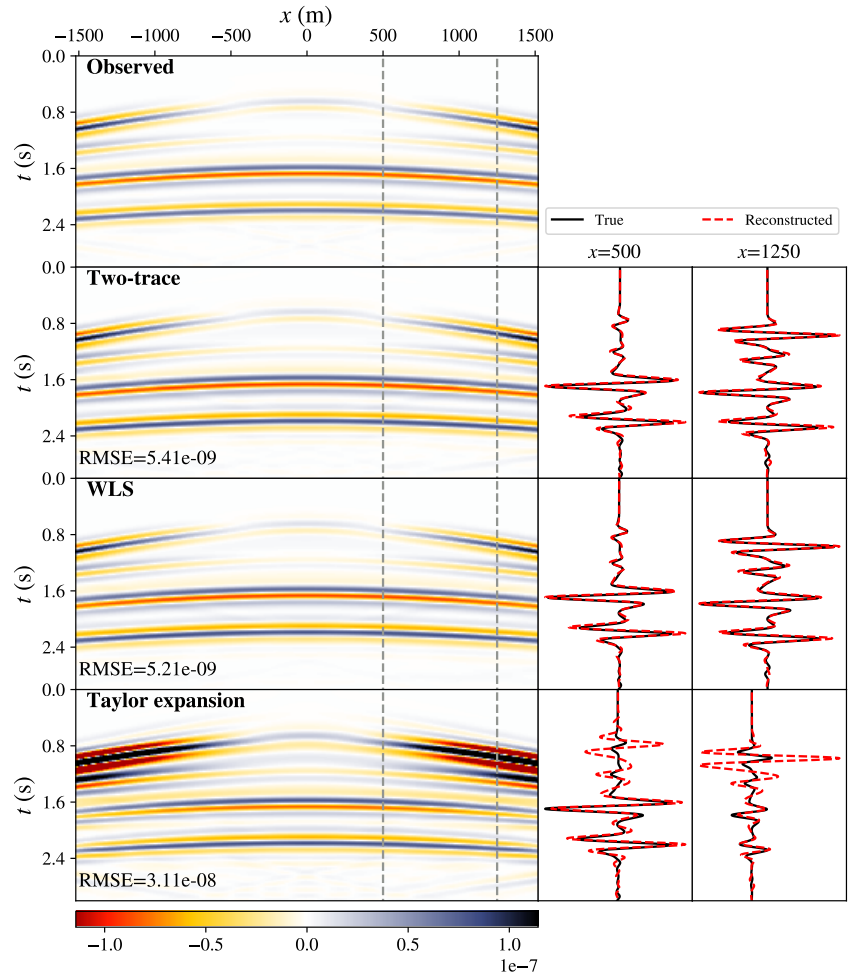
{Left: observed and reconstructed shots obtained via different methods corresponding to correct background models, right: extracted traces. The RMS error between synthetic and observed data is written on each panel. Shots are ploted in the same scale.}

\plot{shots_noRT_short}{width=.9\columnwidth}
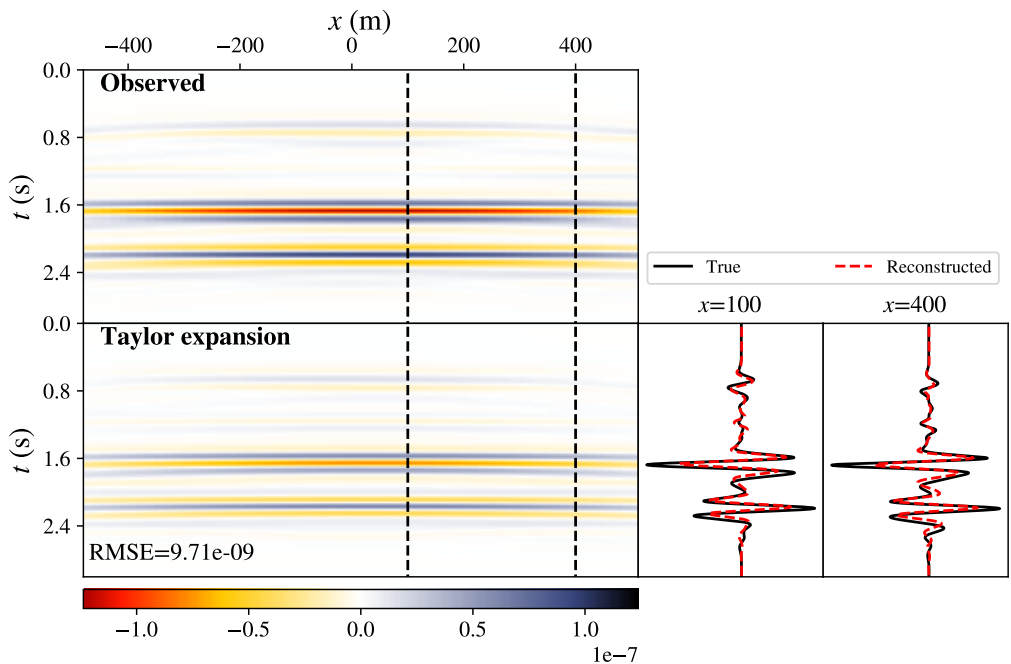
{Left: observed and reconstructed shot obtained via Taylor expansion method corresponding to short-offset acquisition, right: extracted traces.}

To further verify the effectiveness of the different methods, we compare the inverted parameters. First, we invert  the $\beta$ and $\rho$ perturbations via the two-trace method for different angles (Figure~\ref{fig:dif_angle_correct_model1}). As expected, the quality of the inverted parameters may depend on the angle chosen in Equation~\ref{2tracesystem}. To better discriminate between different angles, the RMSE between true and inverted parameters is calculated for each angle (Figure~\ref{fig:dif_angle_error}a). The gray dashed line shows the maximum available angle for the last interface. The error bound for the WLS method is also added to this Figure for comparison. It is seen that the error bound of both parameters for the WLS method is lower than the two-trace method in most of the angles. Moreover, the error for the two-trace method is lower for the mid-range angles comparing to low/high-range angles. We also compare the RMSE in the shot domain for different angles of the two-trace method and the WLS method in Figure~\ref{fig:dif_angle_error}b. As can be seen, the error bound in the shot domain for the WLS method is also lower than the two-trace method in most of the angles. The error bound of the Taylor expansion is not included since it is much more larger than the other methods. We also compare the inversion results for $\beta$ and $\rho$ via the two-trace method ($\gamma'=25^{\circ}$) and the other two methods (Figure~\ref{fig:comparison_correct_model1}). Although both two-trace and WLS methods obtained a good match between inverted and true values, it can be noticed that the WLS method has slightly less oscillations in both inverted $\beta$ and $\rho$ parameters. In contrast, the Taylor expansion inverts the parameters with a lots of oscillations. This is consistent with the fact that weighting the second derivative of $\xi$ in Equation~\ref{taylorRt_t3} with $h^2$, makes the method really sensitive to the artifacts in the CIG panel. 
As expected from parameters cross-talk in radiation pattern (Figure~\ref{fig:radiation}), for the forth event there is a leakage between the two inverted parameters, meaning that the inversion is not unique. In other word, by adding another parameter like density into account, the ill-posedness of the inverse problem is increased since more degree of freedom is considered. Base on the fact that the migrated image mainly comes from the analysis of the reflected waves, and also to reduce the leakage between two parameters, {\color{black} we propose to calculate impedance perturbation ($\delta I_p$) by  non-linear re-parameterization of inverted ${\delta \beta}/{\beta_0}$ and ${\delta \rho}/{\rho_0}$ as \citep{Bharadwaj2018}:
\begin{eqnarray}\label{reparam}
\frac{\delta I_p}{I_{p_0}}  = \sqrt{\frac{\frac{\delta \rho}{\rho_0} + 1}{\frac{\delta\beta}{\beta_0} +1 }} -1,
\end{eqnarray}
for different methods (Figure~\ref{fig:comparison_correct_model1}). The alternative could be to use the linear relationship between impedance, velocity and density perturbations.}  It is seen that the amplitude of the impedance perturbation is much more better estimated comparing to other parameters, meaning that the coupling effect between $\delta \beta$ and $\delta \rho$ is compensated.

Although the system of Equation~\ref{taylorsystem} for the Taylor expansion is well driven, the fact that it is highly sensitive to the maximum surface offset and artifacts in the CIG domain, makes it  inapplicable in sense of seismic application. Consequently, we continue the numerical experiments using the two-trace and the WLS methods.

\plot{dif_angle_correct_model1}{width=1\columnwidth}
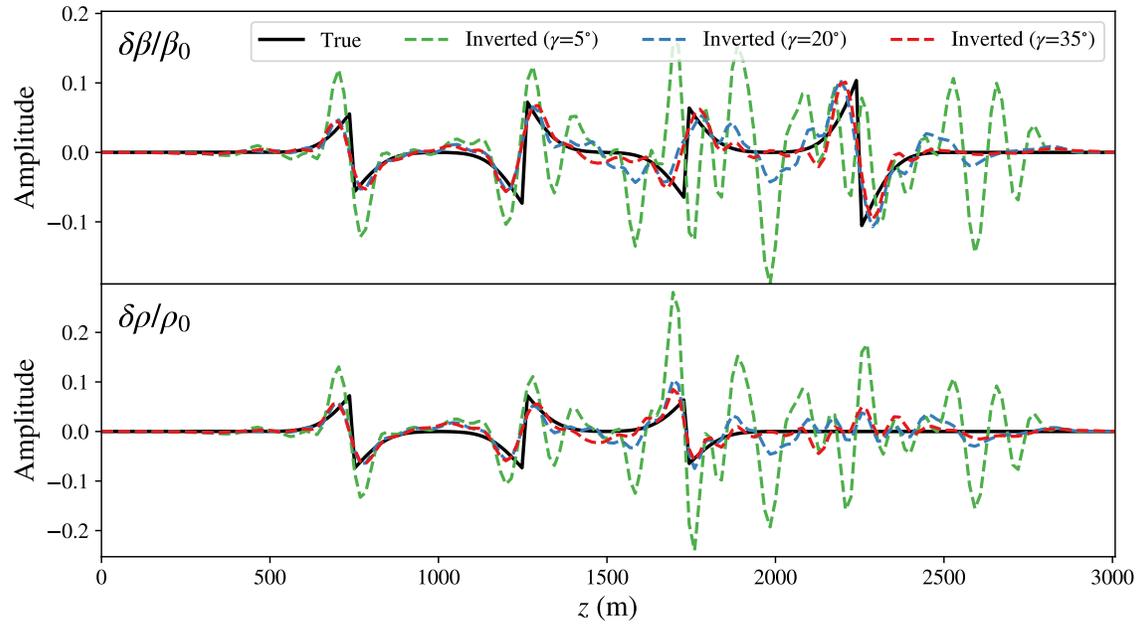
{Inversion results via two-trace method for different angles.}

\plot{dif_angle_error}{width=1\columnwidth}
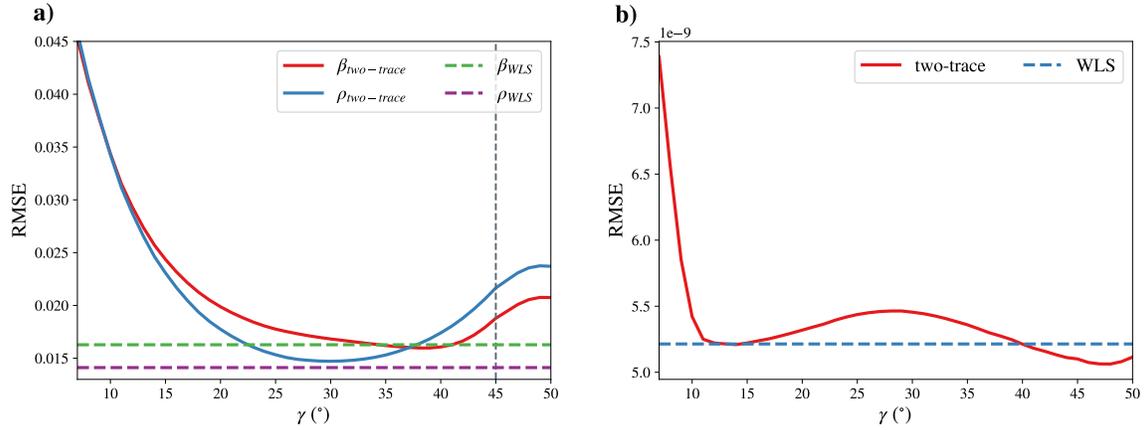
{RMSE corresponding to a) inverted $\beta$ and $\rho$ and b) shot gathers for different angles via two-trace method comparing to WLS method. The gray dashed line in (a) corresponds to maximum available angle for the 4th event.}

\plot{comparison_correct_model1}{width=1\columnwidth}
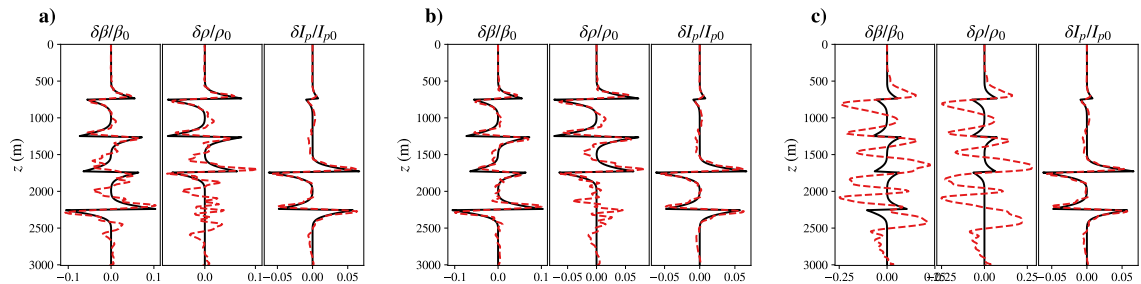
{comparison of inversion results for $\beta$, $\rho$ and $I_p$ obtained via a) two-trace, b) WLS and c) Taylor expansion corresponding to Figure~\ref{fig:cig_correct_model1}. The solid black and dashed red lines correspond to the true and inverted values, respectively.}

\newpage
\subsubsection{\textbf{Sensitivity to background models}}
We now investigate the sensitivity of the methods to background models. We use the same observed data as before, but constant velocity ($v_0=2400$~m/s) and density ($\rho_o=2100$~gr/dm$^3$) models for inversion. The calculated CIG and its Radon transform are shown in Figure~\ref{fig:cig_incorrect_model1}. As expected, an upward curvature is observed because of the too-high velocity for each event. As the CIG contains migration smiles rather than focused points, the integration of Radon transform in CIG domain will not produce the flattened AVA response.  We check the quality of the inversion by comparing computed data and observed data (Figure~\ref{fig:shots_incorrect_model1}). {\color{black} The colorbar scale for reconstructed data is the same as the one for the observed data.} Note that the reconstructed data is modelled via physical inverted parameters in an incorrect model of constant velocity and density. The largest misfit is related to the first event at large offsets. The extracted traces for two-trace method show more misfit comparing to the WLS method in both near and far offsets, specially for the first event. The RMSE between observed and reconstructed shot via each method is also written on each panel showing the superiority of the WLS method. The fact that the remodelled data after inversion nicely matched the observed data in sense of phases and amplitudes, proves that the multiparameter inversion is indeed provides an inverse, even in an incorrect model and modelling in physical domain. {\color{black} Furthermore, this gives some evidences that our method can be coupled to velocity analysis. More research, beyond the scope of this paper, is required to investigate the coupling of multiparameter inversion to velocity analysis.} We also compare the true and the inverted parameters obtained via different methods (Figure~\ref{fig:comparison_incorrect_model1}). As expected, there is a shift in every inverted parameter related to using high-velocity background model.


\plot{cig_incorrect_model1}{width=1\columnwidth}
{Inverted a) $\xi$ and its b) $\mathcal{R}_{\xi}$ at position $x=1500$ m corresponding to Figure~\ref{fig:model_1}.}

\plot{shots_incorrect_model1}{width=.75\columnwidth}
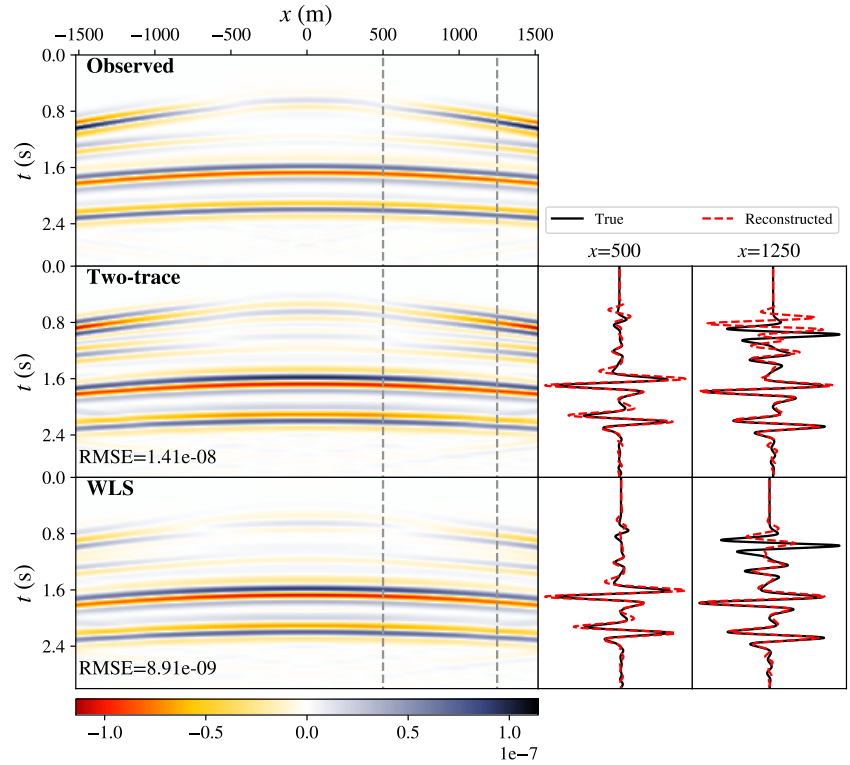
{Left: observed and reconstructed shots obtained via different methods corresponding to incorrect background models, right: extracted traces.}

\plot{comparison_incorrect_model1}{width=1\columnwidth}
{Comparison of inversion results for $\beta$, $\rho$ and $I_p$ obtained via a) two-trace, b) WLS corresponding to Figure~\ref{fig:cig_incorrect_model1}. The solid black and dashed red lines correspond to the true and inverted values, respectively.}

\newpage

\subsection*{Marmousi2 model}
We now consider a more realistic application on the modified Marmousi2 model as a benchmark test. The original Marmousi2 has 17 km width and 3.5 km depth. To reduce the computation cost, we extract the middle part of the Marmousi2 including 10.5 km width and 2.5 km depth, and also to be able to better image the deeper part, we replace the water layer with a higher density layer as in \citet{Yang2016b} and \citet{chen_density}. The exact modified models of velocity (m/s) and density (gr/dm$^3$) of Marmousi2 are shown in Figure~\ref{fig:marmousi_model}. The model is discretized on a $214 \times 875$ grid with $12$ m spacing in $z$ and $x$ direction. The synthetic data correspond to a fixed acquisition geometry (stationary-receivers) with a source spacing of $36$ m and receiver spacing of $12$ m. We use an explosive source, represented by a Ricker wavelet centered at $4.76$ Hz (maximum frequency is $11.9$ Hz). The recording time is 3.7 s, with a time interval of 1.32 ms. 

Here, we investigate the inversion for incorrect and correct background models. As incorrect background models, we use a laterally-homogeneous velocity/density-gradient models (Figure~\ref{fig:marmousi_initial}a-b), and as correct background models, we smooth the true velocity/density models with a 2D Gaussian filter of $60$ m length in both direction (Figure~\ref{fig:marmousi_initial}c-d). The inverted $\xi$ obtained via {\color{black} pseudo-inverse} operator for the incorrect and correct background models at the middle shot (positioned at $5.2$ km) is shown in Figure~\ref{fig:marmousi_cig}a-c, and their angle-domain response via the Radon transform is also shown in Figure~\ref{fig:marmousi_cig}b-d, respectively. The green and black lines is these Figures show the application of the mask for $\alpha=1$ and $\alpha=0.4$ (Figure~\ref{fig:marmousi_initial}b), and $\alpha=0.7$ (Figure~\ref{fig:marmousi_initial}d). First, we compare the data fit in shot domain for the observed shot at position $5.2$ km (Figure~\ref{fig:marmousi_shot}). The reconstructed shots are produced by applying the Born modelling on the physical inverted parameters  obtained via WLS method in the incorrect background models (Figure~\ref{fig:marmousi_shot}b) as well as the correct background models (Figure~\ref{fig:marmousi_shot}c). {\color{black} The colorbar scale for reconstructed data is the same as the one for the observed data.} To evaluate the reliability, the extracted traces for different offsets are also shown in Figure~\ref{fig:marmousi_shot}. It is seen that the shot corresponding to incorrect background models has a correct phase and satisfactory amplitudes match, specially for the short offset and shallow part. A detailed look at the incorrect velocity model (Figure~\ref{fig:marmousi_initial}a) and comparing it with correct one (Figure~\ref{fig:marmousi_initial}c), implies that almost in all depths and locations higher velocities are considered as an incorrect model comparing to the correct velocity model. Similar to the results in Figure~\ref{fig:comparison_incorrect_model1}, this leads to a shift and defocusing in the inverted parameters. Therefore, the error in the deeper part is expected as the acquisition depth is not enough. For the correct background models, once more, there is a almost perfect match in sense of the phase and amplitude. Obviously, we would also expect small differences because of the geologic complexity of the model, which can be reduced by further LSM iterations. 

We also compare the result of inverted physical parameters. The exact perturbation models and the inverted perturbation models for incorrect and correct background models are respectively shown in the first, second and third row of Figure~\ref{fig:marmousi_perturbationt}. As mentioned before, it is evident that there is shift in the inverted parameters for the incorrect background models (Figure~\ref{fig:marmousi_perturbationt}d-f). The extracted traces at different positions for further comparison between the exact and inverted parameters in correct background models are shown in Figure~\ref{fig:marmousi_perturbation_extract}. Although the phase information is accurately preserved, the amplitudes of the $\delta \beta$ and $\delta \rho$ have small differences and show a small leakage between two parameters. Nevertheless, the amplitudes of the $\delta I_p$ is well recovered, as for the former test.

These different numerical tests prove that the inversion formulas developed for the 2D variable density acoustic media are indeed inverse instead of adjoint: almost perfect fit to the observed data even in incorrect background models. Moreover, the proposed method shows better and more robust results comparing to the one proposed by \citet{Dafni2018}, in both image and shot domains. More interestingly, the results for impedance perturbation obtained by combination of other inverted parameters, is in better accordance with the true one and mitigates the ill-posedness of the inverse problem.

\plot{marmousi_model}{width=.95\columnwidth}
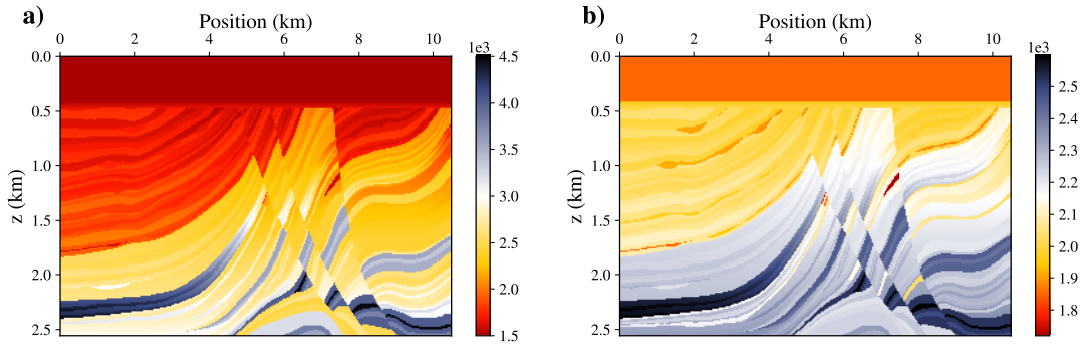
{The exact a) velocity (m/s) and b) density (g/cm$^3$) corresponding to the Marmousi2 Model.}

\plot{marmousi_initial}{width=.95\columnwidth}
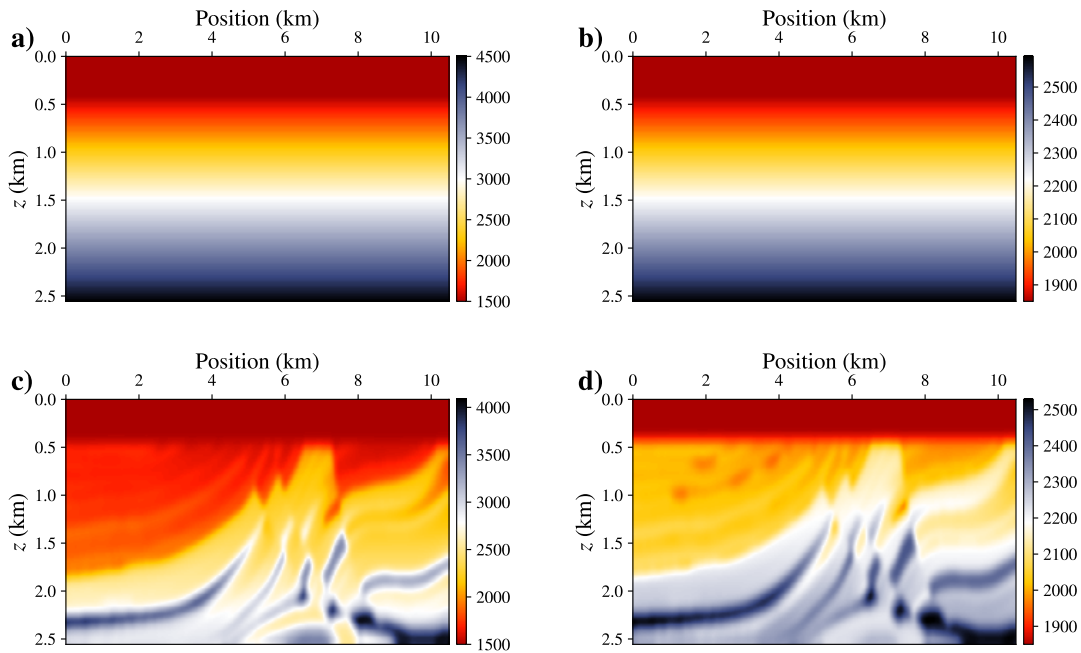
{The initial incorrect background a) velocity and b) density model and the initial correct background c) velocity and d) density model.}

\plot{marmousi_cig}{width=.95\columnwidth}
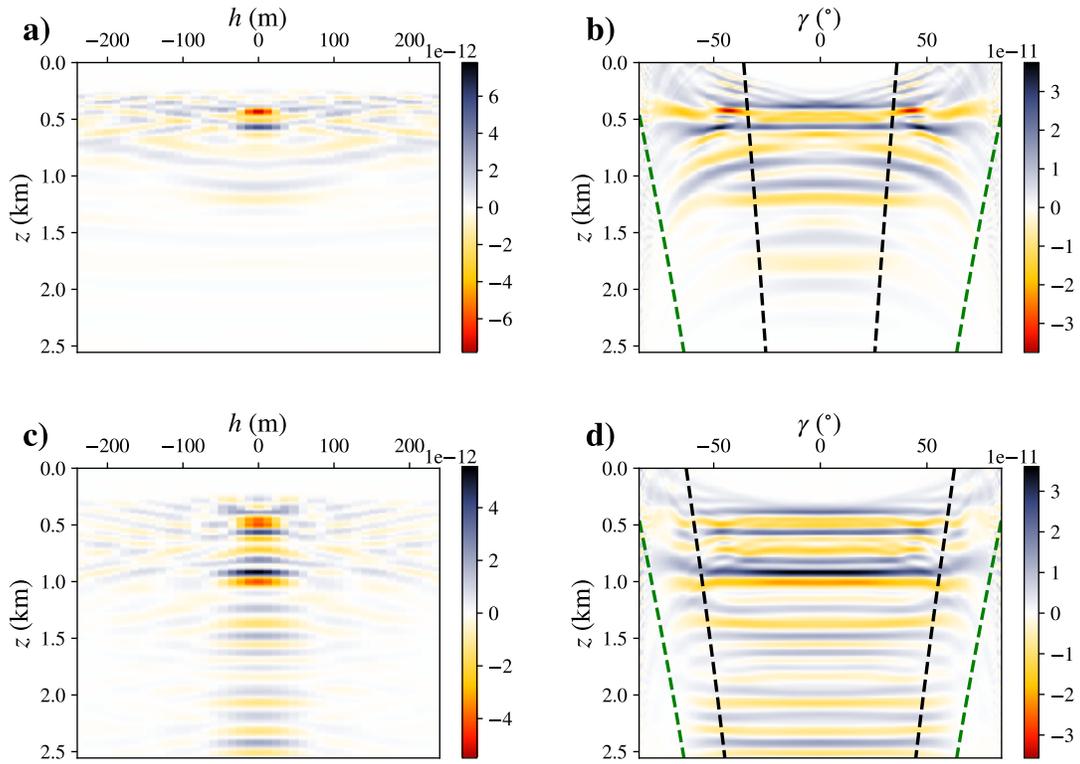
{Inverted a) $\xi$ and its b) $\mathcal{R}_{\xi}$ corresponding to incorrect background model and inverted c) $\xi$ and its d) $\mathcal{R}_{\xi}$ corresponding to correct background model at position $x=5.2$~km. The green and black line in (b) and (d) corresponds to the limit for acquired angles and the used ones in inversion, respectively.}

\plot{marmousi_shot}{width=.95\columnwidth}
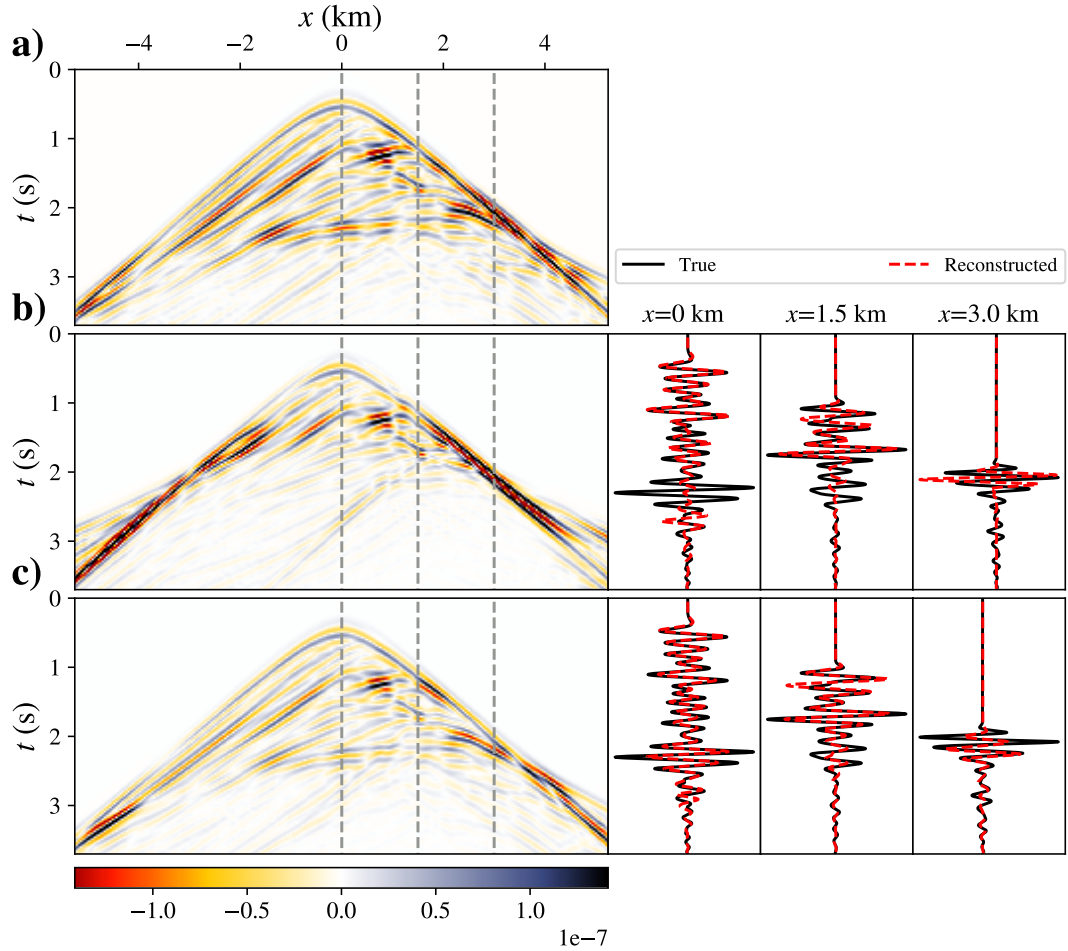
{Shot gathers of a) observed data and reconstructed data for the shot at position $5.2$ km with b) incorrect background models and c) correct background models. The extracted traces at different offsets are shown in left of each shot.}

\plot{marmousi_perturbationt}{width=1\columnwidth}
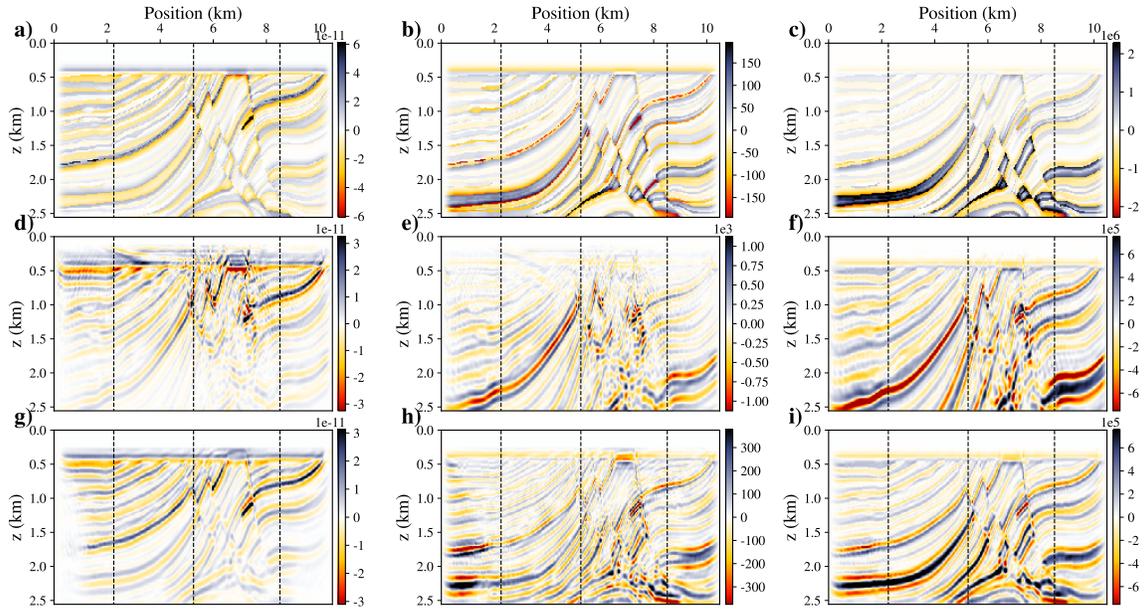
{The exact perturbation model for a) $\beta$, b) $\rho$ and c) $I_p$, and the inverted perturbation model for d) $\beta$, e) $\rho$ and f) $I_p$ corresponding to incorrect background models and g) $\beta$, h) $\rho$ and i) $I_p$ corresponding to correct background models. The dashed line corresponds to extracted traces.}

\plot{marmousi_perturbation_extract}{width=1\columnwidth}
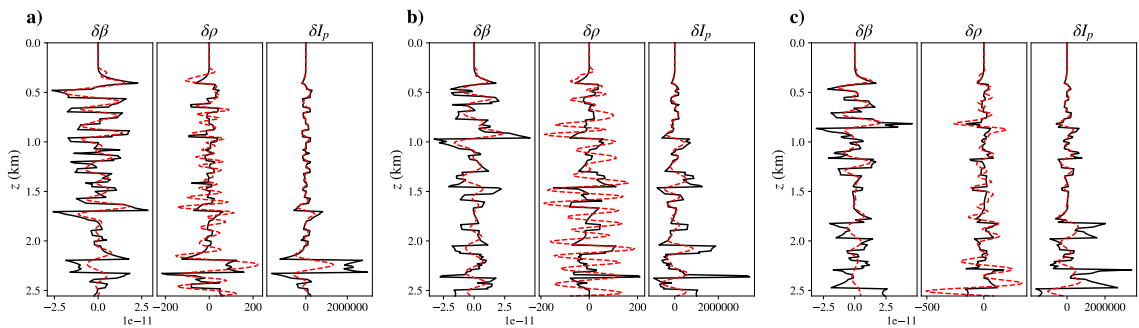
{The extracted traces from perturbation models (Figure~\ref{fig:marmousi_perturbationt}) for a) 2.2~km, b) 5.2~km and c) 8.5~km. The solid black and dashed red lines correspond to the true and
inverted values, respectively.}

\section{Discussion}
We proposed two new approaches by generalization of the two-trace method as well as the Taylor expansion of the Radon transform. The latter was able to reconstruct the observed data only for short-offset acquisition. This is too limiting in practice, as seismic imaging requires to handle long-offset data. Consequently, the proposed method in this paper is the WLS method, which had better results for both correct and incorrect background models. An interesting property of such a generalization is that it gives us the necessary flexibility to add the regularization term to the objective function (Equation~\ref{J3}), leading to constrained optimization problem. The objective is to mitigate the ill-posedness resulting from parameters cross-talk based on the radiation pattern (Figure~\ref{fig:radiation}) or to add a priori information. 

Recently, \citet{Qin2016} proposed an approach to jointly invert the velocity and density in preserved-amplitude Full Waveform Inversion (FWI). The strategies developed here have two main differences with the one proposed by \citet{Qin2016}. First, our inversion is based on the extended domain (subsurface offset), which is not the case in \citet{Qin2016}. Second and more importantly, we calculate the diffraction angle ($\gamma$) either with the Radon transform or the Taylor expansion, whereas they calculate it by applying the tomographic ray tracing, which is not necessarily consistent with the wave equation-based approach. In practice, they apply an iterative process to possibly reduce the approximation errors introduced in the estimation of the angles. These make the methods proposed here more consistent to the one proposed by \citet{Qin2016}.

We investigate the effect of the choice of the parameterization in the inversion. We decompose the resulted $\xi_\beta$ of the simple model (Figure~\ref{fig:cig_correct_model1}) to two different parameters based on the corresponding diffraction patterns (Table~\ref{parametrization}) in Equation~\ref{J3} and \ref{J_mat} (Figure~\ref{fig:wls_reparametrization}). The third parameter in each subplot of Figure~\ref{fig:wls_reparametrization} is inferred from the combination of the first two parameters. The good agreement between the results of different parameterization suggests that the choice of parameterization does not change the final results in our work. The conclusion differs in the case of FWI. In non-linear imaging approaches such as FWI, the choice of the parameterization is not neutral, meaning that the final results depend on parameterization class \citep{Tarantola1984,prieux2013}, whereas migration, a linear operator by definition, does not suffer from this aspect.  

\begin{table}[h]\caption{The different diffraction patterns for different parameterization classes.}
\centering
\renewcommand{\arraystretch}{1.2}
\begin{tabular}{ccc}
\Cline{2pt}{1-3}
Parameterization 	&First parameter	& Second parameter   \\
\hline
$(\beta,\rho)$		& $-1$				& $\cos(2\gamma)$	     \\
$(I_p,\rho)$		& $2$				& $-2\sin^2(\gamma)$	  \\
$(V_p,\rho)$		& $2$				& $2\cos^2(\gamma)$	      \\
$(V_p,I_p)$			& $2\sin^2(\gamma)$		& $2\cos^2(\gamma)$	   \\
\hline
\end{tabular}\label{parametrization}
\end{table}

\plot{wls_reparametrization}{width=1\columnwidth}
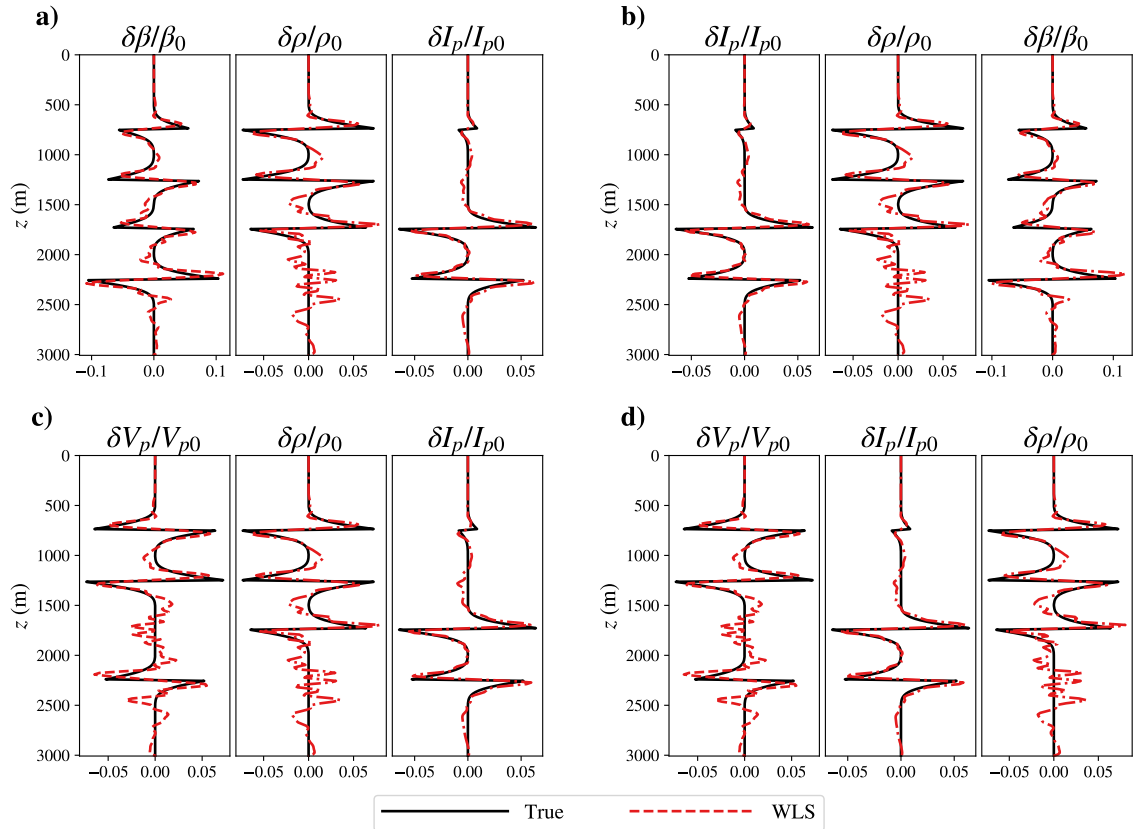
{The inverted parameters for different parameterization classes as a) $(\beta,\rho)$, b) $(I_p,\rho)$, c) $(V_p,\rho)$ and d) $(V_p,I_p)$. The third parameter in each panel is inferred by combination of the first two parameters. }

In the case of incorrect background model, a mismatch of data fit in shallow part was observed for the simple model (specially first event), whereas it was observed in the deeper part for  Marmousi2. The reasons for these observations are different. In the case of the simple model, the incorrect background models are constant velocity/density models which are very different from the correct background models, leading to extremely defocused energy in inverted $\xi$ (Figure~\ref{fig:cig_incorrect_model1}a). Meanwhile, the incorrect background models for Marmousi2 are laterally-homogeneous velocity/density-gradient models which are closer to the correct background models. This is also noticeable by comparing the defocused energy of $\xi$ for different models (Figure~\ref{fig:cig_incorrect_model1}a and \ref{fig:marmousi_cig}c). Accordingly, the shallow part in Marmousi2 will have better data fit as the incorrect background model is close to correct one, which is not the case in simple model. {\color{black} In order to analyze the effect of the subsurface offset for the mismatch of data fit in the deeper part of the Marmousi2, we run an additional test by doubling this parameter (Figure~\ref{fig:marmousi_rev_discus}). It can be noticed that the mismatch in the deeper part of the reconstructed shot (Figure~\ref{fig:marmousi_rev_discus}c) is more less the same as before (Figure~\ref{fig:marmousi_shot}b), meaning that this error is not caused by the truncation in the CIG domain}. As already has been mentioned in the numerical experiments, using higher background velocity model leads to a downward shift in the inverted reflectivity (Figure~\ref{fig:comparison_incorrect_model1} and \ref{fig:marmousi_perturbationt}). Thus, if the recording depth is not enough to image the reflectivity, a mismatch in the deeper part of the reconstructed shot would indeed be expected, which is the case for the Marmousi2. 

\plot{marmousi_rev_discus}{width=.9\columnwidth}
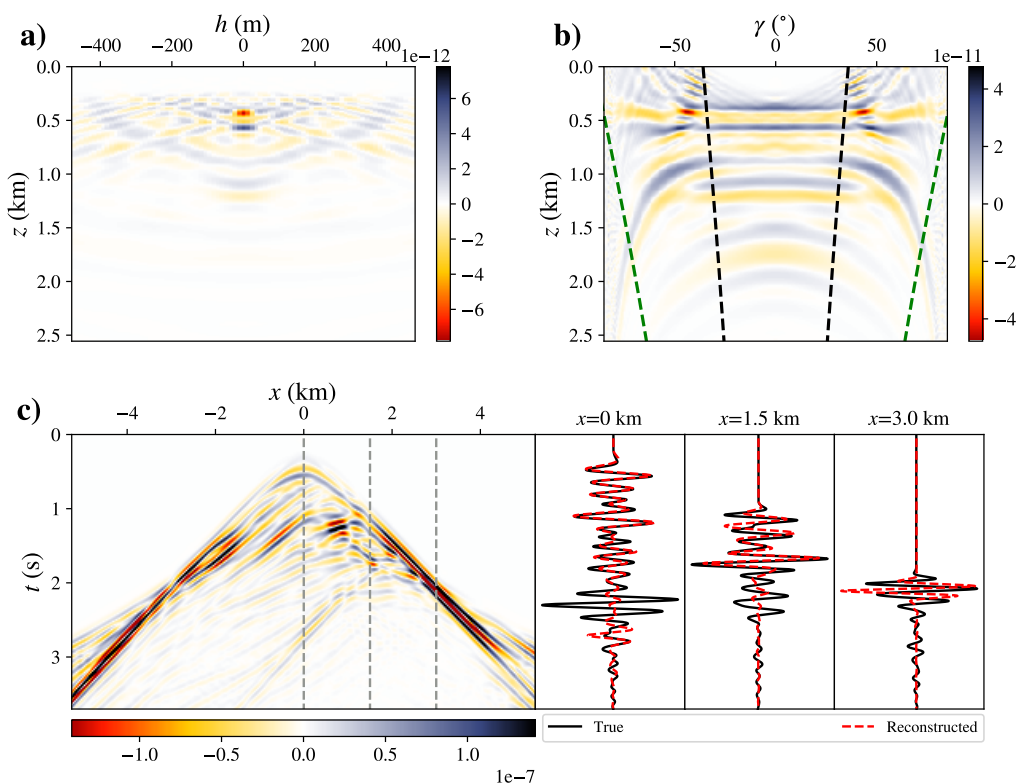
{Inverted a) $\xi$ and its b) $\mathcal{R}_{\xi}$ corresponding to incorrect background model and 480 m extension of subsurface offset. c) Reconstructed data for the shot at position $5.2$ km with incorrect background models. The extracted traces at different offsets are shown in left on the shot gather.}

In term of implementation, the variable density {\color{black} pseudo-inverse} Born modelling consists of two operators, the {\color{black} pseudo-inverse} Born modelling and the forward Radon transform operators. The Radon transform has a long history of application in seismic processing, for instance, velocity analysis \citep{Thorson1985}, multiple attenuation \citep{Hampson1986}, NMO-free stacking \citep{Gholami_decon} and AVO-preserved processing \citep{Farshad_avo,Gholami_shuey}. Based on the path of integration in the Radon transform, many effective methods for rapid evaluation of the traditional Radon transform has been proposed \citep{Hu_butterfly,Nikitin2017,Gholami_gfst,Gholami_chirp1,Gholami_chirp2}. Here, since the size of each slice in CIG domain is constant, it is possible to explicitly construct the matrix for the Radon transformation. The application of this matrix for the Radon transformation has a computational complexity of $O(N_x N_z N_h N_{\gamma})$, whereas wave-equation based operators have computational complexity of $O(N_x N_z N_s N_t)$, given that the numbers of samples for $z$, $h$, $x$, $\gamma$, $t$ and sources are $N_z$, $N_h$, $N_x$, $N_{\gamma}$, $N_t$ and $N_s$. Note that $N_h$ samples cross-correlation should be also performed between the calculated source and receiver wavefields. Several techniques such as computing the CIG only at a specific image points \citep{Yang2015} or computing the CIG with only a random choice of traces \citep{vanLeeuwen} have been proposed to reduce the computational burden of cross-correlation, but not the propagation. Besides, the value of $N_s \times N_t$ is much larger than $N_h \times N_{\gamma}$ in practice. Therefore, the main computational burden of the variable density {\color{black} pseudo-inverse} Born modelling is due to the modelling operators and it remains at the same order as that of the constant density. It is also worth to note that there is no need to apply the inverse of the Radon transform here. Otherwise, the cost would be different as it should be solved iteratively. 

\section{Conclusions}
In this paper, we have proposed an efficient weighted least-squares approach to extend the constant density {\color{black} pseudo-inverse} Born modelling to variable density acoustic media. This is a generalization of the method proposed by \citet{Dafni2018}. It is {\color{black} here} based on using the whole AVA response in the angle domain. We have also proposed another approach based on the Taylor expansion of the Radon transform, which does not require application of the Radon transform. Numerical experiments proves that the {\color{black} latter} is not applicable in sense of seismic imaging {\color{black} because of the noise enhancement at large subsurface offset}, whereas the weighted least-squares method is very promising and provides robust results when compared with the other approaches. We conclude that such a generalization also provides the flexibility to include more constraints in inversion. Future work will consist of including regularization terms in the least-squares objective function, {\color{black} coupling multiparameter inversion to velocity analysis,} and also extending the {\color{black} pseudo-inverse} Born modelling operator beyond the acoustic case.

\newpage
\bibliographystyle{seg}
\bibliography{paper}

\begin{thebibliography}{}
\itemsep0pt

\bibitem[Baysal et~al., 1983]{rtm1983}
Baysal, E., D.~D. Kosloff, and J.~W. Sherwood,  1983, Reverse time migration:
  Geophysics, {\bf 48}, 1514--1524.

\bibitem[Bednar, 2005]{migbednar}
Bednar, J.~B.,  2005, A brief history of seismic migration: Geophysics, {\bf
  70}, 3MJ--20MJ.

\bibitem[Beylkin, 1985]{Beylkin1985}
Beylkin, G.,  1985, Imaging of discontinuities in the inverse scattering
  problem by inversion of a causal generalized {R}adon transform: Journal of
  Mathematical Physics, {\bf 26}, 99--108.

\bibitem[Bharadwaj et~al., 2018]{Bharadwaj2018}
Bharadwaj, P., W. Mulder, and G. Drijkoningen,  2018, A parameterization
  analysis for acoustic full--waveform inversion of sub--wavelength anomalies.

\bibitem[Biondi and Symes, 2004]{Biondi2004}
Biondi, B., and W.~W. Symes,  2004, Angle-‐domain common‐-image gathers for
  migration velocity analysis by wavefield-‐continuation imaging: Geophysics,
  {\bf 69}, 1283--1298.

\bibitem[Bleistein, 1987]{Bleistein1987}
Bleistein, N.,  1987, On the imaging of reflectors in the {E}arth: Geophysics,
  {\bf 52}, 931--942.

\bibitem[Brandsberg‐Dahl et~al., 2003]{Brandsberg}
Brandsberg‐Dahl, S., M.~V. de~Hoop, and B. Ursin,  2003, Focusing in dip and
  {AVA} compensation on scattering--angle/azimuth common image gathers:
  Geophysics, {\bf 68}, 232--254.

\bibitem[Castagna and Smith, 1994]{Castagna1994}
Castagna, J.~P., and S.~W. Smith,  1994, Comparison of {AVO} indicators: A
  modeling study: Geophysics, {\bf 59}, 1849--1855.

\bibitem[Chauris and Cocher, 2017]{chauris2017}
Chauris, H., and E. Cocher,  2017, From migration to inversion velocity
  analysis: Geophysics, {\bf 82}, S207--S223.

\bibitem[Chauris and Cocher, 2018]{chauris2018EAGE}
--------, 2018, Review of different expressions for the extended born
  approximate inverse operator: Presented at the 80th EAGE Conference \&
  Exhibition 2018 Workshop Programme.

\bibitem[Chen and Sacchi, 2018]{chen_density}
Chen, K., and M.~D. Sacchi,  2018, Should we include the density perturbation
  in elastic least-squares reverse time migration?, {\it in} SEG Technical
  Program Expanded Abstracts 2018: Society of Exploration Geophysicists,
  4226--4230.

\bibitem[Chopra and Castagna, 2014]{avo2014}
Chopra, S., and J.~P. Castagna,  2014, {AVO}: Society of Exploration
  Geophysicists.

\bibitem[Claerbout, 1971]{Claerbout1971}
Claerbout, J.~F.,  1971, Toward a unified theory of reflector mapping:
  Geophysics, {\bf 36}, 467--481.

\bibitem[Claerbout, 1985]{Claerbout1985}
--------, 1985, Imaging the {E}arth's interior: Blackwell scientific
  publications Oxford.

\bibitem[Dafni and Symes, 2016]{Dafni2016}
Dafni, R., and W.~W. Symes,  2016, Scattering and dip angle decomposition based
  on subsurface offset extended wave--equation migration: Geophysics, {\bf 81},
  S119--S138.

\bibitem[Dafni and Symes, 2018]{Dafni2018}
--------, 2018, Asymptotic inversion of the variable density acoustic model,
  {\it in} SEG Technical Program Expanded Abstracts 2018: Society of
  Exploration Geophysicists,  570--574.

\bibitem[Dai et~al., 2012]{Dai_LSRTM2012}
Dai, W., P. Fowler, and G.~T. Schuster,  2012, Multi--source least-squares
  reverse time migration: Geophysical Prospecting, {\bf 60}, 681--695.

\bibitem[Dai et~al., 2011]{Dai_LSRTM2011}
Dai, W., X. Wang, and G.~T. Schuster,  2011, Least--squares migration of
  multisource data with a deblurring filter: Geophysics, {\bf 76}, R135--R146.

\bibitem[de~Bruin et~al., 1990]{debruin}
de~Bruin, C. G.~M., C.~P.~A. Wapenaar, and A.~J. Berkhout,  1990,
  Angle--dependent reflectivity by means of prestack migration: Geophysics,
  {\bf 55}, 1223--1234.

\bibitem[de~Hoop and Bleistein, 1997]{GRT1997}
de~Hoop, M.~V., and N. Bleistein,  1997, Generalized {R}adon transform
  inversions for reflectivity in anisotropic elastic media: Inverse Problems,
  {\bf 13}, 669--690.

\bibitem[Dutta and Schuster, 2014]{Dutta2014Attenuation}
Dutta, G., and G.~T. Schuster,  2014, Attenuation compensation for
  least--squares reverse time migration using the viscoacoustic--wave equation:
  Geophysics, {\bf 79}, S251--S262.

\bibitem[Farshad et~al., 2018]{Farshad_avo}
Farshad, M., A. Gholami, and A.~M. Mobarakeh,  2018, {AVO}--preserving
  deconvolutive {R}adon transform: Presented at the 80th EAGE Conference and
  Exhibition 2018.

\bibitem[Forgues and Lambar{\'e}, 1997]{Forgues}
Forgues, E., and G. Lambar{\'e},  1997, Parameterization study for acoustic and
  elastic ray + {B}orn inversion: Journal of Seismic Exploration, {\bf 6},
  253--277.

\bibitem[Gholami, 2017]{Gholami_decon}
Gholami, A.,  2017, Deconvolutive {R}adon transform: Geophysics, {\bf 82},
  V117--V125.

\bibitem[Gholami and Farshad, 2019a]{Gholami_chirp2}
Gholami, A., and M. Farshad,  2019a, Fast hyperbolic {R}adon transform using
  chirp--z transform: Digital Signal Processing, {\bf 87}, 34--42.

\bibitem[Gholami and Farshad, 2019b]{Gholami_shuey}
--------, 2019b, The {S}huey--{R}adon transform: Geophysics, {\bf 84},
  V197--V206.

\bibitem[Gholami and Sacchi, 2017]{Gholami_gfst}
Gholami, A., and M.~D. Sacchi,  2017, Time--invariant {R}adon transform by
  generalized {F}ourier slice theorem: Inverse Problems \& Imaging, {\bf 11}.

\bibitem[Gholami and Zand, 2017]{Gholami_chirp1}
Gholami, A., and T. Zand,  2017, Fast $\ell_1$--regularized {R}adon transforms
  for seismic data processing: Digital Signal Processing, {\bf 71}, 83--94.

\bibitem[Hampson, 1986]{Hampson1986}
Hampson, D.,  1986, Inverse velocity stacking for multiple elimination, {\it
  in} SEG Technical Program Expanded Abstracts 1986: Society of Exploration
  Geophysicists,  422--424.

\bibitem[Hou and Symes, 2015]{symes2015}
Hou, J., and W.~W. Symes,  2015, An approximate inverse to the extended {B}orn
  modeling operator: Geophysics, {\bf 80}, R331--R349.

\bibitem[Hou and Symes, 2017]{symes2017a}
--------, 2017, An alternative formula for approximate extended {B}orn
  inversion: Geophysics, {\bf 82}, S1--S8.

\bibitem[Hou and Symes, 2018]{symes2017b}
--------, 2018, Inversion velocity analysis in the subsurface--offset domain:
  Geophysics, {\bf 83}, R189--R200.

\bibitem[Hu et~al., 2013]{Hu_butterfly}
Hu, J., S. Fomel, L. Demanet, and L. Ying,  2013, A fast butterfly algorithm
  for generalized {R}adon transforms: Geophysics, {\bf 78}, U41--U51.

\bibitem[Lailly and Bednar, 1983]{Lailly1983}
Lailly, P., and J. Bednar,  1983, The seismic inverse problem as a sequence of
  before stack migrations: Conference on inverse scattering: theory and
  application, Siam Philadelphia, PA, 206--220.

\bibitem[LeBras and Clayton, 1988]{Lebras_LSM}
LeBras, R., and R.~W. Clayton,  1988, An iterative inversion of back--scattered
  acoustic waves: Geophysics, {\bf 53}, 501--508.

\bibitem[Li and Chauris, 2018]{Yubing2018}
Li, Y., and H. Chauris,  2018, Coupling direct inversion to common--shot
  image--domain velocity analysis: Geophysics, {\bf 83}, R497--R514.

\bibitem[Montel and Lambar\'e, 2011]{Montel2011angle}
Montel, J., and G. Lambar\'e,  2011, {RTM} and {K}irchhoff angle domain
  common--image gathers for migration velocity analysis, {\it in} SEG Technical
  Program Expanded Abstracts 2011:  3120--3124.

\bibitem[Mulder and Plessix, 2004]{rtm_comparison}
Mulder, W.~A., and R. Plessix,  2004, A comparison between one--way and
  two--way wave--equation migration: Geophysics, {\bf 69}, 1491--1504.

\bibitem[Nemeth et~al., 1999]{Nmeth_LMS}
Nemeth, T., C. Wu, and G.~T. Schuster,  1999, Least--squares migration of
  incomplete reflection data: Geophysics, {\bf 64}, 208--221.

\bibitem[Nikitin et~al., 2017]{Nikitin2017}
Nikitin, V.~V., F. Andersson, M. Carlsson, and A.~A. Duchkov,  2017, Fast
  hyperbolic {R}adon transform represented as convolutions in log--polar
  coordinates: Computers \& Geosciences, {\bf 105}, 21 -- 33.

\bibitem[Operto et~al., 2013]{Operto2013parametrization}
Operto, S., Y. Gholami, V. Prieux, A. Ribodetti, R. Brossier, L. Metivier, and
  J. Virieux,  2013, A guided tour of multiparameter full--waveform inversion
  with multicomponent data: From theory to practice: The Leading Edge, {\bf
  32}, 1040--1054.

\bibitem[Prieux et~al., 2013]{prieux2013}
Prieux, V., R. Brossier, S. Operto, and J. Virieux,  2013, {Multiparameter full
  waveform inversion of multicomponent ocean--bottom--cable data from the
  Valhall field. Part 1: imaging compressional wave speed, density and
  attenuation}: Geophysical Journal International, {\bf 194}, 1640--1664.

\bibitem[Qin and Lambar\'e, 2016]{Qin2016}
Qin, B., and G. Lambar\'e,  2016, Joint inversion of velocity and density in
  preserved--amplitude full--waveform inversion, {\it in} SEG Technical Program
  Expanded Abstracts 2016: Society of Exploration Geophysicists,  1325--1330.

\bibitem[Sava and Alkhalifah, 2013]{Sava2013}
Sava, P., and T. Alkhalifah,  2013, Wide--azimuth angle gathers for anisotropic
  wave--equation migration: Geophysical Prospecting, {\bf 61}, 75--91.

\bibitem[Sava and Fomel, 2006]{sava2006extent}
Sava, P., and S. Fomel,  2006, Time--shift imaging condition in seismic
  migration: Geophysics, {\bf 71}, S209--S217.

\bibitem[Sava and Vlad, 2011]{Sava2011}
Sava, P., and I. Vlad,  2011, Wide--azimuth angle gathers for wave--equation
  migration: Geophysics, {\bf 76}, S131--S141.

\bibitem[Sava and Fomel, 2003]{sava2003RT}
Sava, P.~C., and S. Fomel,  2003, Angle--domain common--image gathers by
  wavefield continuation methods: Geophysics, {\bf 68}, 1065--1074.

\bibitem[Sun et~al., 2018]{sun2018ELSRTM}
Sun, M., L. Dong, J. Yang, C. Huang, and Y. Liu,  2018, Elastic least--squares
  reverse time migration with density variations: Geophysics, {\bf 83},
  S533--S547.

\bibitem[Symes, 2008]{Symes2008}
Symes, W.~W.,  2008, Migration velocity analysis and waveform inversion:
  Geophysical Prospecting, {\bf 56}, 765--790.

\bibitem[Tarantola, 1984]{Tarantola1984}
Tarantola, A.,  1984, Inversion of seismic reflection data in the acoustic
  approximation: Geophysics, {\bf 49}, 1259--1266.

\bibitem[ten Kroode, 2012]{tenkroode}
ten Kroode, F.,  2012, A wave--equation-based {K}irchhoff operator: Inverse
  Problems, {\bf 28}, 115013.

\bibitem[Thorson and Claerbout, 1985]{Thorson1985}
Thorson, J.~R., and J.~F. Claerbout,  1985, Velocity--stack and slant--stack
  stochastic inversion: Geophysics, {\bf 50}, 2727--2741.

\bibitem[van Leeuwen et~al., 2015]{vanLeeuwen}
van Leeuwen, T., R. Kumar, and F. Herrmann,  2015, Affordable full subsurface
  image volume--an application to wemva: Presented at the 77th EAGE Conference
  and Exhibition-Workshops.

\bibitem[Virieux and Operto, 2009]{virieuxfwi}
Virieux, J., and S. Operto,  2009, An overview of full--waveform inversion in
  exploration geophysics: Geophysics, {\bf 74}, WCC1--WCC26.

\bibitem[Xue et~al., 2016]{Xue_LSRTM2016}
Xue, Z., Y. Chen, S. Fomel, and J. Sun,  2016, Seismic imaging of incomplete
  data and simultaneous--source data using least--squares reverse time
  migration with shaping regularization: Geophysics, {\bf 81}, S11--S20.

\bibitem[Yang et~al., 2016a]{Yang2016b}
Yang, J., Y. Liu, and L. Dong,  2016a, Least--squares reverse time migration in
  the presence of density variations: Geophysics, {\bf 81}, S497--S509.

\bibitem[Yang et~al., 2016b]{Yang2016}
--------, 2016b, Simultaneous estimation of velocity and density in acoustic
  multiparameter full--waveform inversion using an improved
  scattering--integral approach: Geophysics, {\bf 81}, R399--R415.

\bibitem[Yang and Sava, 2015]{Yang2015}
Yang, T., and P. Sava,  2015, Image--domain wavefield tomography with extended
  common--image--point gathers: Geophysical Prospecting, {\bf 63}, 1086--1096.

\bibitem[Zeng et~al., 2014]{Zeng_LSRTM}
Zeng, C., S. Dong, and B. Wang,  2014, Least--squares reverse time migration:
  {I}nversion-based imaging toward true reflectivity: The Leading Edge, {\bf
  33}, 962--968.

\bibitem[Zhang and Schuster, 2014]{Zhang_LSRTM2014}
Zhang, D., and G.~T. Schuster,  2014, Least--squares reverse time migration of
  multiples: Geophysics, {\bf 79}, S11--S21.

\bibitem[Zhang et~al., 2015]{Zhang_LSRTM2015}
Zhang, Y., L. Duan, and Y. Xie,  2015, A stable and practical implementation of
  least--squares reverse time migration: Geophysics, {\bf 80}, V23--V31.

\bibitem[Zhang et~al., 2014]{Zhang_MultiRTM2014}
Zhang, Y., A. Ratcliffe, G. Roberts, and L. Duan,  2014, Amplitude--preserving
  reverse time migration: {F}rom reflectivity to velocity and impedance
  inversion: Geophysics, {\bf 79}, S271--S283.

\bibitem[Zhang et~al., 2005]{Zhang_LSM2005}
Zhang, Y., G. Zhang, and N. Bleistein,  2005, Theory of true--amplitude
  one--way wave equations and true--amplitude common--shot migration:
  Geophysics, {\bf 70}, E1--E10.

\end{thebibliography}

\end{document}